\documentclass{aa}
\usepackage{booktabs}
\usepackage{natbib}
\usepackage{hyperref}
\usepackage{footnote}
\usepackage{tablefootnote}
\usepackage{graphicx}
\usepackage{txfonts}
\usepackage{longtable}
\usepackage{url}
\usepackage{float,color}
\usepackage{subcaption}
\usepackage{multirow}

\hypersetup{
    colorlinks=true,
    linkcolor=blue,
    filecolor=magenta,      
    urlcolor=cyan,
    citecolor=blue
}

\newcommand{\Elow}{E_{\rm low}}
\newcommand{\Eup}{E_{\rm up}}

\newcommand{\td}{\langle {\rm 3D} \rangle}

\newcommand{\stagger}{STAGGER}
\newcommand{\cobold}{CO5BOLD}
\newcommand{\Ot}[5]{\mbox{$#1\,^#2{\rm #3}^{{\rm #4}}_{\rm #5}$}}

\definecolor{esgreen}{RGB}{15, 98, 18}

\begin{document} 

\title{Observational constraints on the origin of the elements. IV: \\
The standard composition of the Sun}
\subtitle{}

\author{
   Ekaterina Magg \inst{\ref{inst:MPIA}},
   Maria Bergemann \inst{\ref{inst:MPIA}, \ref{inst:NBIA}},
   Aldo Serenelli \inst{\ref{inst:ICE},\ref{inst:IEEC},\ref{inst:MPIA}},
   Manuel Bautista \inst{\ref{inst:Bautista}},
   Bertrand Plez \inst{\ref{inst:montp}}, 
   Ulrike Heiter \inst{\ref{inst:Uppsala}},
   Jeffrey M. Gerber \inst{\ref{inst:MPIA}},
   Hans-G\"unter Ludwig \inst{\ref{inst:lsw}}, 
   Sarbani Basu
   \inst{\ref{inst:yale}}, 
    Jason W. Ferguson \inst{\ref{inst:wichita}}
   Helena Carvajal Gallego \inst{\ref{inst:umons}},
   S\'ebastien Gamrath \inst{\ref{inst:umons}},
   Patrick Palmeri \inst{\ref{inst:umons}},
   Pascal Quinet \inst{\ref{inst:umons}, \ref{inst:ulg}}
 }

\institute{
Max Planck Institute for Astronomy, K\"{o}nigstuhl 17, 69117, Heidelberg, Germany \email{emagg@mpia-hd.mpg.de}
\label{inst:MPIA}
\and
Institute of Space Sciences (ICE, CSIC), Carrer de Can Magrans S/N, E-08193, Cerdanyola del Valles, Spain
\label{inst:ICE}
\and
Institut d'Estudis Espacials de Catalunya (IEEC), Carrer Gran Capita 2, E-08034, Barcelona, Spain
\label{inst:IEEC}
\and
Department of Physics, Western Michigan University, Kalamazoo, Michigan, 49008, USA
\label{inst:Bautista}
\and
Niels Bohr International Academy, Niels Bohr Institute, University of Copenhagen
Blegdamsvej 17, DK-2100 Copenhagen, Denmark
\email{bergemann@mpia-hd.mpg.de}
\label{inst:NBIA}
\and
Observational Astrophysics, Department of Physics and Astronomy, Uppsala University, Box 516, 75120 Uppsala, Sweden
\label{inst:Uppsala}
\and
LUPM, Univ Montpellier, CNRS, Montpellier, France
\label{inst:montp}
\and
Landessternwarte -- Zentrum f\"ur Astronomie der Universit\"at
Heidelberg, K\"{o}nigstuhl 12, 69117 Heidelberg, Germany
\label{inst:lsw}
\and
Department of Astronomy, Yale University, PO Box 208101, New Haven, CT 06520-8101, USA
\label{inst:yale}
\and
Physics Department, Wichita State University, Wichita, KS 67260-0032, USA
\label{inst:wichita}
\and
Physique Atomique et Astrophysique, Universit\'e de Mons, B-7000 Mons, Belgium
\label{inst:umons}
\and
IPNAS, Universit\'e de Li\`ege, B-4000 Li\`ege, Belgium
\label{inst:ulg}
}

\date{Received ; accepted }

\abstract{The chemical composition of the Sun is requested in the context of various studies in astrophysics, among them in the calculation of the standard solar models (SSMs), which describe the evolution of the Sun from the pre-main-sequence to its present age.}
{In this work, we provide a critical re-analysis of the solar chemical abundances and corresponding SSMs.}
{For the photospheric values, we employ new high-quality solar observational data collected with the IAG facility, state-of-the art non-equilibrium modelling, new oscillator strengths, and different atmospheric models, including the MARCS model, but also averages based on Stagger and CO5BOLD 3D radiation-hydrodynamics simulations of stellar convection. We perform new calculations of oscillator strengths for transitions in O I and N I. For O I - the critical element for the interior models - calculations are carried out using several independent methods. We find unprecedented agreement between the new estimates of transition probabilities, thus supporting our revised solar oxygen abundance. We also provide new estimates of the noble gas Ne abundance.}
{We investigate our results in comparison with the previous estimates. We discuss the consistency of our photospheric measurements with meteoritic values taking into account systematic and correlated errors. Finally, we provide revised chemical abundances, leading to a new value of the solar photospheric present-day metallicity $Z/X = 0.0225$, and employ them in the calculations of the SSM. We find that the puzzling mismatch between the helioseismic constraints on the solar interior structure and the model is resolved with the new chemical composition.}
{}

\keywords{sun: abundances -- stars: atmospheres -- atomic data -- line: formation -- radiative transfer}
\titlerunning{The standard composition of the Sun}
\authorrunning{E. Magg, et al.}
\maketitle
%
\section{Introduction}
Research in modern astrophysics shows an increasingly growing interest in high-precision stellar abundance diagnostics, as a source of accurate knowledge of the chemical composition of stars is relevant in studies of exoplanets \citep[e.g.][]{Bedell2018, Adibekyan2019}, asteroseismology and stellar structure \citep{Nissen2017, Deal2020}, and Galaxy evolution \citep{Bensby2014,Bergemann2018,Schuler2021}. Chemical abundances can only be determined from stellar spectra, and therefore, with the Sun being the reference for any chemical diagnostics study, this effort requires self-consistent unbiased analyses of stellar and solar data. However, so far, accurate studies of solar abundances \citep{Caffau2011,Asplund2021} adopted methods and data, which are conceptually different from those applied to large stellar samples. In particular, solar abundances are usually determined using full three-dimensional (3D) radiation transfer not assuming local thermodynamic equilibrium (NLTE) methods employing spatially-resolved solar spectra taken at different pointings across the solar disc \citep[e.g.][]{Amarsi2018,Bergemann2021}. Also, typically very weak atomic and molecular features across the entire range from the optical to mid-IR at $\sim 1.5~\mu$m are used for the solar analysis (such as the lines of OH, CN, CH, and NH, \citealt{Amarsi2021}), which is usually inapplicable for large samples of stars. In addition, full 3D NLTE calculations are computationally prohibitive and are currently not feasible for large stellar samples. Most large spectroscopic surveys, such as Gaia-ESO \citep{Smiljanic2014}, APOGEE \citep{Majewski2017}, and RAVE \citep{Steinmetz2020} still have to rely on spectroscopic models computed under simplifying assumptions of local thermodynamic equilibrium (LTE), one-dimensional (1D) geometry, and hydrostastic equilibrium, adopting parameterizations for convective energy transport and turbulence. GALAH \citep{2021MNRAS.506..150B} 
is the only survey that has adopted NLTE grids so far. Also, spatially-resolved spectra are not available for any other star than the Sun. Next-generation astronomical facilities, such as 4MOST and WEAVE, have stringent requirements on the quality of chemical abundance characterisation, but they will rely on medium resolution spectral data.

\begin{figure}[h!]
\includegraphics[width=1.0\linewidth]{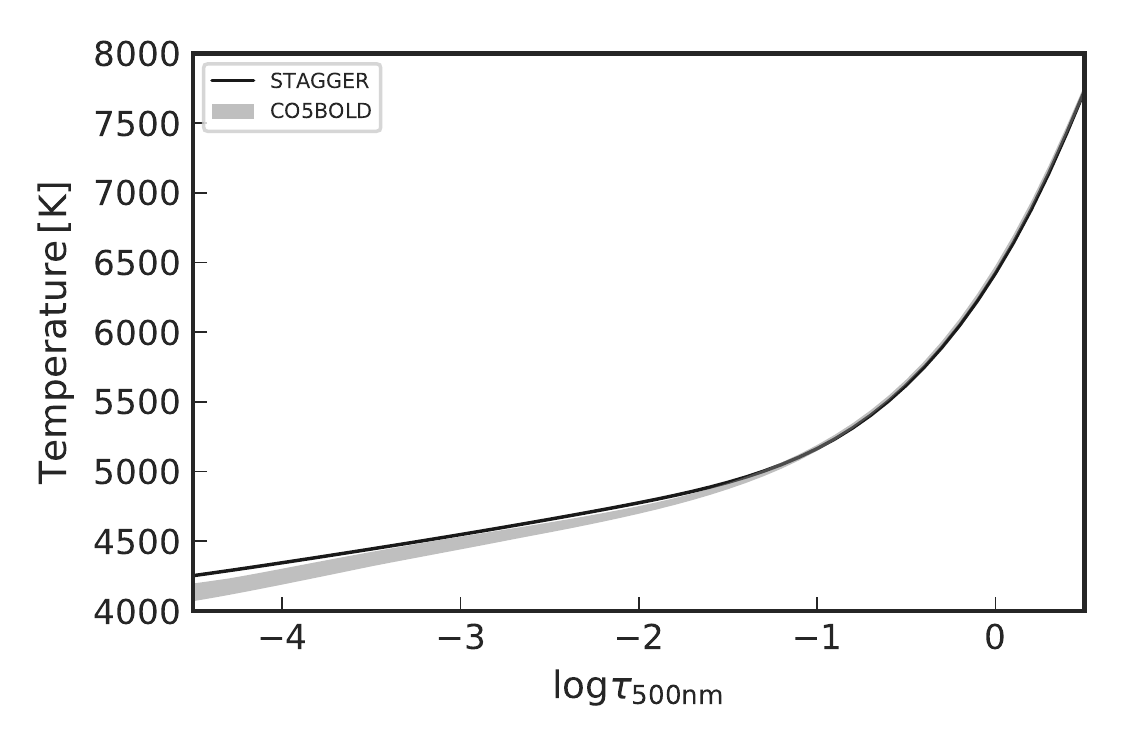}
\includegraphics[width=1.0\linewidth]{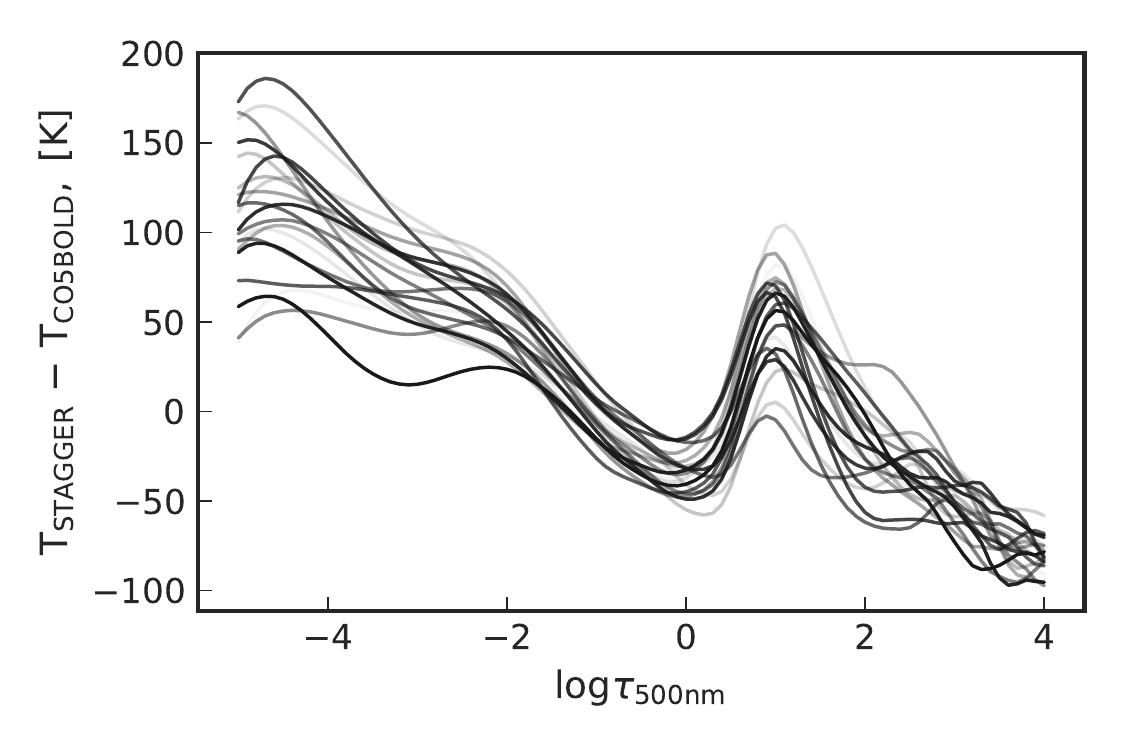}
\caption{ \textit{Top panel:} temperature structure of \stagger\ and \cobold\ model atmospheres. Gray area depicts variations among temporarily resolved snapshots in \cobold\, simulations.
\textit{Bottom panel:} absolute difference in temperature structure of \stagger\ and \cobold\ model atmospheres. Each line corresponds to a temporarily resolved snapshot of \cobold\ simulations.
See Sect.~\ref{sec:stagger_cobold_comp} for details.
}
\label{fig:stagger_cobold_temperature}
\end{figure} 
In this work, we provide a new analysis of the solar chemical composition.
The methodology is chosen such that it is suitable for the analysis of any star, not only the Sun with its very high quality observations.
We use the most up-to-date atomic and molecular data \citep[and updates described in Sect.~\ref{sec:linelists}]{Heiter2021}, new NLTE model atoms \citep[e.g.][]{Semenova2020, Bergemann2021}, and different solar model atmospheres (CO5BOLD and STAGGER) obtained by averaging 3D radiation-hydrodynamics (RHD) simulations of stellar convection. The latter is important in the view of the debate over the solar abundances between the two groups who use the STAGGER and CO5BOLD models. 
As the full 3D RHD computations are not applicable for the up-coming large spectroscopic surveys, given the complexity of the wide wavelength range abundance analysis, for this study we choose to work with 1D and average 3D model atmospheres.
We focus primarily on those chemical elements that are relevant in the calculation of the standard solar models (SSMs), that is C, N, O, Mg, Si, Ca Fe, and Ni. We also carefully revisit various observational constraints on the abundance of Ne, which cannot be determined from the solar photospheric spectra. However, independent measurements based on the solar wind and corona are available \citep{Bochsler2007}. We select  atomic spectral lines that can also be used to derive abundances from medium resolution spectral data.
We also discuss the photospheric measurements of O \citep[e.g.][]{Bochsler2007,Laming2017} in the context of Ne/O ratio. We compare our results with previous estimates in the literature, also with those based on the analysis of the B-type stars in the solar neighbourhood \citep{Nieva2011,Nieva2012}.

The paper is organized as follows. In Sect.~\ref{sec:analysis} we describe the observational material and the solar model atmospheres employed in this work. We summarize the key details of NLTE model atoms, statistical equilibrium and synthetic spectra calculations, and the input line list. In Sect.~\ref{sec:abundances}, we present our new solar abundance estimates and compare them with the literature. We close with the analysis of the new chemical composition in the SSM calculations in Sect.~\ref{sec:ssm} and draw conclusions in Sect.~\ref{sec:conclusions}. 
\section{Analysis}\label{sec:analysis}
\subsection{Observed data}\label{sec:observations}
We use a high-quality, high-resolution ($R = \lambda/\Delta\lambda \approx 700\,000$) solar flux spectrum obtained with the FTS instrument at the Institut f\"ur Astrophysik, Göttingen \citep[hereafter, IAG data]{IAG2016}. 
In \citet{Bergemann2021}, we investigated the differences between solar abundance estimates obtained using the IAG data, the KPNO FTS solar flux atlas, the data acquired with the \textit{Hinode} space based facility, and several other datasets. We found that in some cases, non-negligible differences arise due to the use of different solar atlases \citep[see also][]{Caffau2008}. However, the differences are primarily associated with instrumental artefacts and effects of the data reduction. As a consequence, part of the discrepancies between the solar abundance estimates (e.g. \citealt{Asplund2021} and \citealt{Caffau2011}) arguably arise because of the latter aspect, since they focus on weak features that are particularly sensitive to the details of the continuum placement.

In this work, we do not restrict the analysis to the weakest lines, but also include other spectral lines that will be accessible with next generation facilities such as 4MOST and WEAVE.  As we will show in Sect.~\ref{sec:abundances}, we do not detect any significant systematic biases in abundances, caused by using lines across a broad range of equivalent widths, as long as the atomic data employed in the analysis are of a sufficiently high quality. 
\subsection{Model atmospheres}\label{sec:atmos}

We make use of three different sources of solar atmospheric models: the MARCS model \citep{Gustafsson2008}, the STAGGER model \citep{Magic2013a,Magic2013b}, and the CO5BOLD model \citep{Freytag2012}. 

The physical properties of the MARCS model atmosphere code were extensively described in \citet{Gustafsson2008}. This is a 1D LTE model atmosphere computed under the assumption of hydrostatic equilibrium and with convective energy transfer treated according to the mixing length theory formalism of \citet{Henyey1965}. The mixing length was set to $1.5$ and microturbulence to $1$ km\,s$^{-1}$. The code relies on opacities computed using the Uppsala opacity package, which was updated to include comprehensive bound-bound and bound-free transitions in all relevant absorbers in FGK atmospheres, as well as lines  for about $20$ important molecular species. In total over $500$ molecules are included in the equation of state calculations. 

Here, we use the averages of these simulations\footnote{\url{https://staggergrid.wordpress.com/mean-3d/}}, constructed by spatial (over surfaces of equal optical depth) and temporal averaging of the simulation cubes, as described in \citet{Bergemann2012}. These averages are known as $\td$ model atmospheres.
To represent the original 3D velocity field in the simulation cubes we include a depth-dependent velocity profile in the form of a microturbulence with a value of one standard deviation of the 3D velocity components as suggested in \citet{Uitenbroek2011}.
The opacities and the equation of state used in the RHD calculations are described in \citet{Magic2013a,Magic2013b}.

The mean \cobold\ model \citep[see][for a description of the code]{Freytag+al12} was obtained by horizontal (on $\tau_\mathrm{Rosseland}$ iso-surfaces, first temperature moment) averaging of data blocks from a solar model simulation (internal identifier {\tt d3gt57g44msc600}). The model is part of the ongoing extensive efforts for the development of the {\tt CIFIST} 3D model atmosphere grid \citep{Ludwig+al09,Tremblay+al13}. The 3D model uses $250 \times 250 \times 207$ grid points, with an equidistant grid spacing of 32\,km in the two horizontal directions, and a non-equidistant grid spacing in the vertical direction between 10 and 15\,km, giving a total extension of $8.0\times 8.0\times 2.3$\,Mm$^3$. The wavelength dependence of the radiative transfer was represented with $12$ opacity bins.
A comparison of the temperature structures of the STAGGER and CO5BOLD model atmospheres used in this study is presented in Fig.~\ref{fig:stagger_cobold_temperature}. 
\begin{figure}
\includegraphics[width=1.0\linewidth]{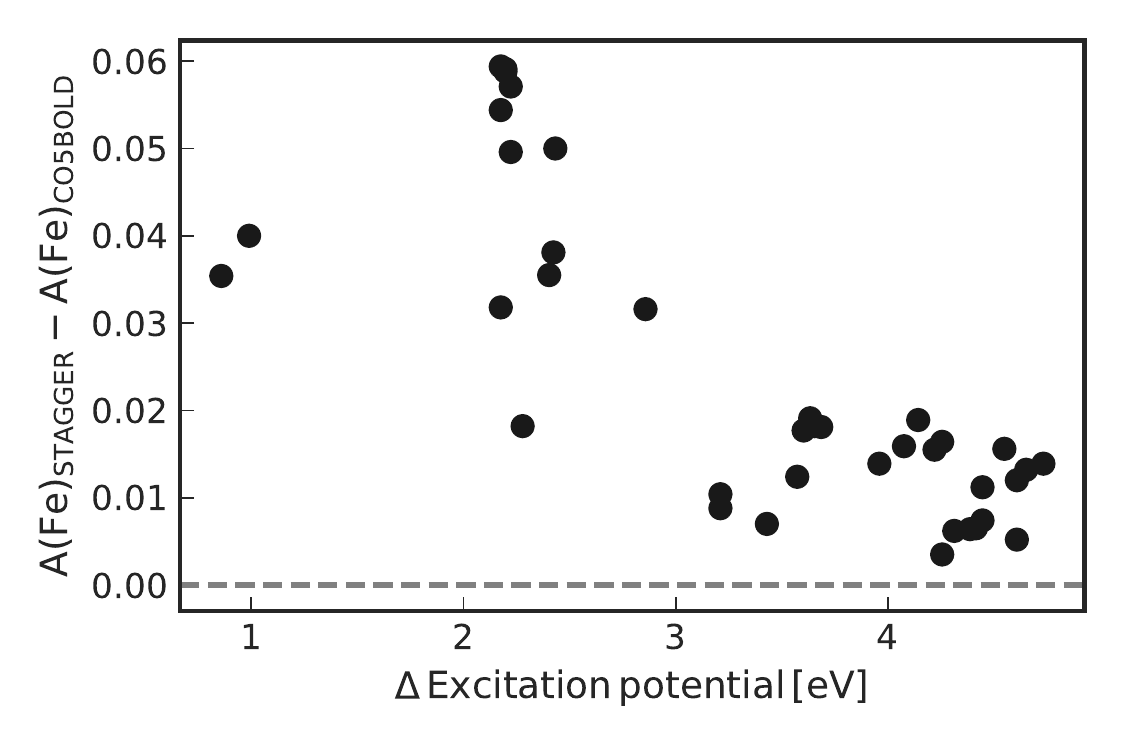}
\includegraphics[width=1.0\linewidth]{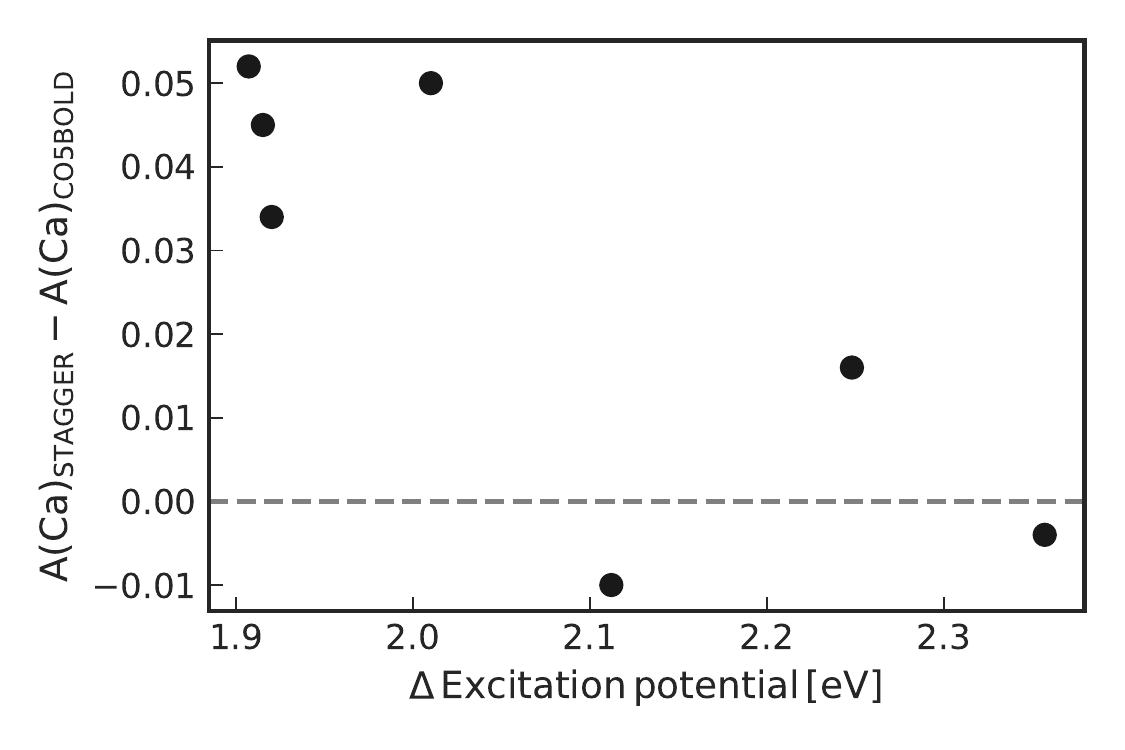}
\caption{\textit{Top panel:} predicted difference in Fe abundance computed with \stagger\ and \cobold\ model atmospheres as a function of lower level excitation potential for Fe I lines. \textit{Bottom panel:} predicted difference in Ca abundance computed with \stagger\ and \cobold\ model atmospheres as a function of difference in lower level excitation potential for Ca I lines. See Sect.~\ref{sec:stagger_cobold_comp} for details.}
\label{fig:stagger_cobold_abundances}
\end{figure} 
\subsection{Comparative analysis of \cobold\ and \stagger\ results}
\label{sec:stagger_cobold_comp}

To test how the choice of model atmosphere affects the derived abundances, we computed the curve-of-growth (COG) for each diagnostic line of each chemical element using the \cobold\, and \stagger\, solar atmospheric models. Since there is no consensus yet on how to average the 3D velocity field, we chose to include velocity in a form of microturbulence and we set it to $1$ km\,s$^{-1}$ in both model atmospheres. By comparing the COGs, we estimated the difference in abundance for each chemical element. It is known that NLTE radiative transfer is much less sensitive to the temperature structure of the model atmosphere \citet{Bergemann2012}, because the populations are significantly affected by the non-local radiation field. Thus, we perform the comparison in NLTE whenever possible. This predicted  difference is provided in Table~\ref{table:1d3d_abundances} and we refer to it later in the discussion when comparing abundances derived using the \cobold\ and \stagger~ models.

Overall, the \stagger\ model atmosphere is hotter than \cobold\ (Fig. \ref{fig:stagger_cobold_temperature}), with a negligible difference around the optical depth $\log{\tau_{\mathrm{500nm}}}=0$, but up to $150$~K at larger optical depths. 
Fig.~\ref{fig:stagger_cobold_abundances} shows that Fe I lines with lower level excitation potential $\Elow \lesssim 2.5$~eV are very sensitive to the temperature structure of the atmosphere, and the abundances inferred from these features may differ by up to $+0.06$ dex depending on the detailed structure of the $\td$ model. In contrast, the Fe abundances derived from Fe I lines with higher $\Elow$ values change by less than $0.02$ dex. This is as expected, and confirms previous NLTE results  \citep[e.g.][]{Bergemann2012}. Therefore, in this work, we have chosen to include only Fe I lines with high excitation potential, $\Elow \geq 2.5$~eV in the abundance analysis.

Also, the lines of Ca I show a significant sensitivity to the structure of the models. The majority of diagnostic Ca I lines in our linelist arise from levels with intermediate excitation potential, $\Elow \approx 2.5$~eV. 
As Fig.~\ref{fig:stagger_cobold_abundances} demonstrates, the strong Ca I lines at 6471 and 6499 \AA\ (with the smallest energy level differences) have the largest difference in abundance when modelled with the \cobold\ and \stagger\ model, up to $0.05$ dex.
Since we do not currently have a  robust evidence in favour of either of these 3D RHD simulations, we opt to include all Ca lines in the  analysis, given the much smaller number of diagnostic features of Ca I available for the 
abundance analysis compared to Fe I.
The lines of all other elements, see Table~\ref{table:1d3d_abundances}, appear to be almost unaffected by the differences between \cobold\, and \stagger\, models. Therefore we use all available diagnostic lines in the calculations, not imposing any cuts on their atomic parameters.
\subsection{Abundance analysis}\label{sec:models}
The analysis of spectral lines is a two-step procedure. First, we compute NLTE atomic level populations for each of the NLTE elements described in Sec. \ref{sec:nlteatoms} using a code that simultaneously solves the radiative transfer and statistical equilibrium equations. Second, the NLTE populations are used in a spectrum synthesis code to compute model spectra for all diagnostic lines of the selected chemical elements, taking all atomic and molecular blends self-consistently into account. The abundances are then derived by comparing the grid of model spectra with the data, employing standard $\chi^2$ minimisation. 
We have verified on mock lower-resolution observed data that the abundances are not affected by reducing the resolving power and signal-to-noise ratio of the spectra by more than 0.03 dex.

For the NLTE calculations, we use the MULTI2.3 code \citep{Carlsson1986}, which is based on the method of accelerated lambda iteration (or ALI) and solves the radiative transfer equation using the long characteristics solver. The code was updated, as described in \citet{Bergemann2019} and \citet{Gallagher2020}.
The atomic populations calculated by MULTI2.3 are then used with the spectrum synthesis code Turbospectrum \citep{Plez2012}, that can self-consistently treat blends and compute a full spectrum including all species simultaneously.
\begin{figure*}[h!]
\hbox{
\includegraphics[width=0.5\linewidth]{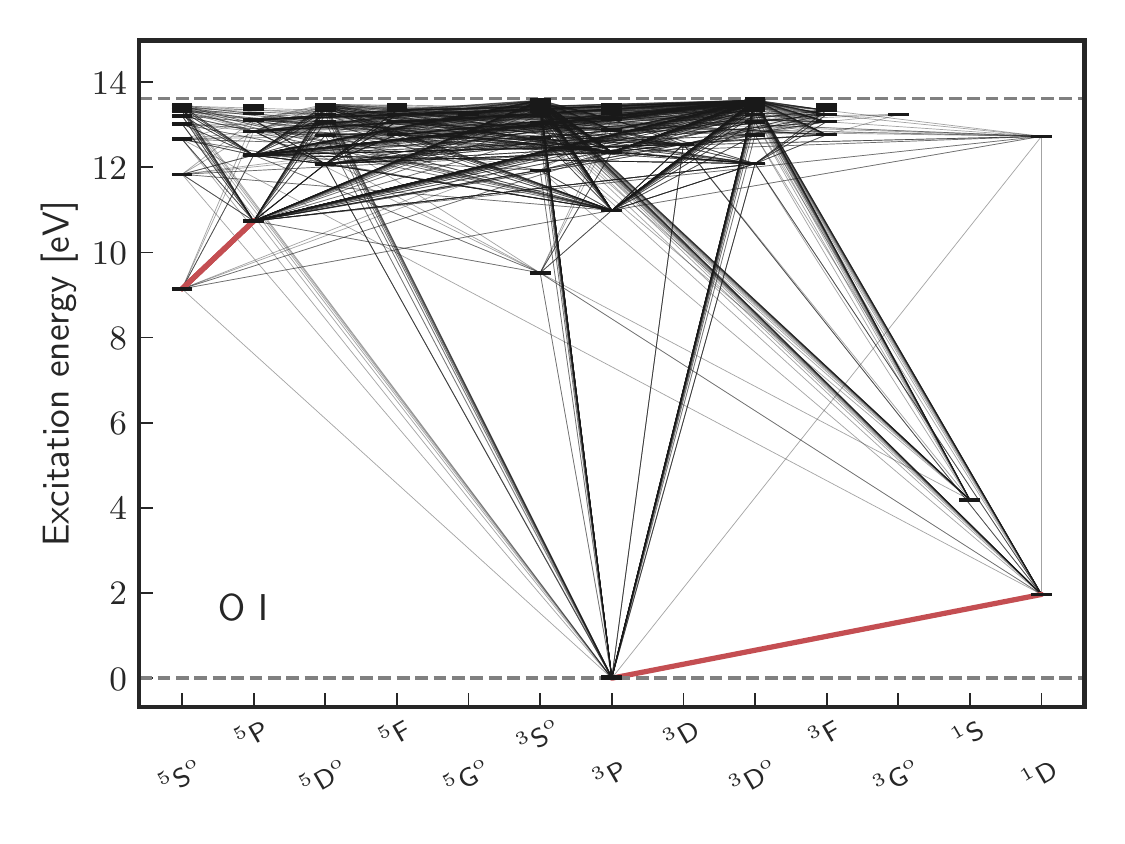}
\includegraphics[width=0.5\linewidth]{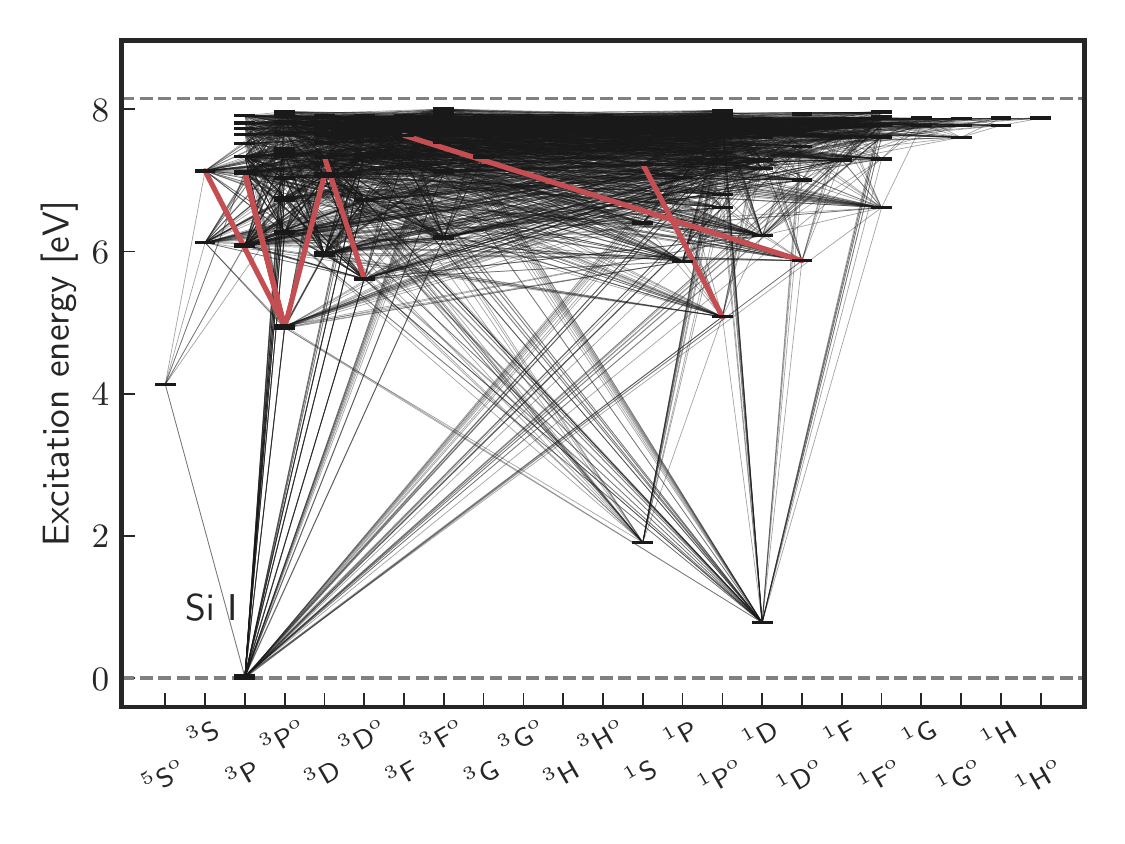}
}
\hbox{
\includegraphics[width=0.5\linewidth]{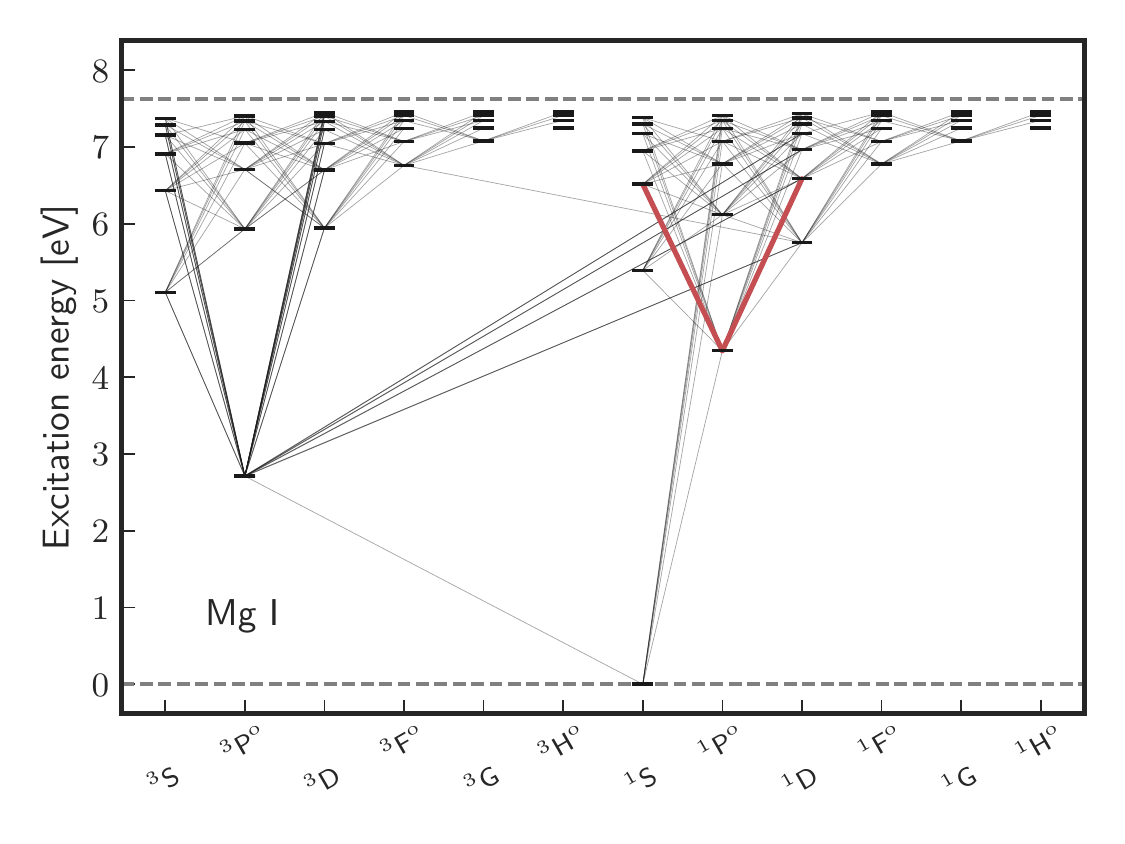}
\includegraphics[width=0.5\linewidth]{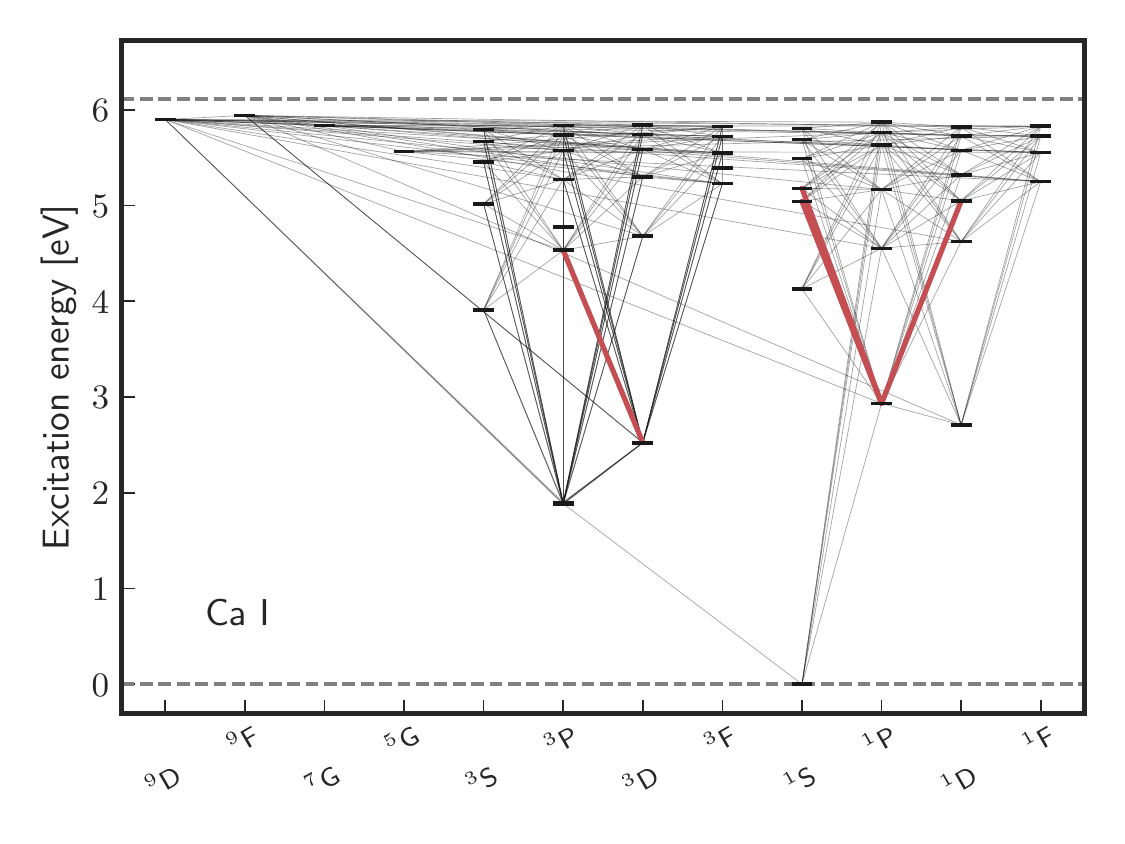}
}
\hbox{
\includegraphics[width=0.5\linewidth]{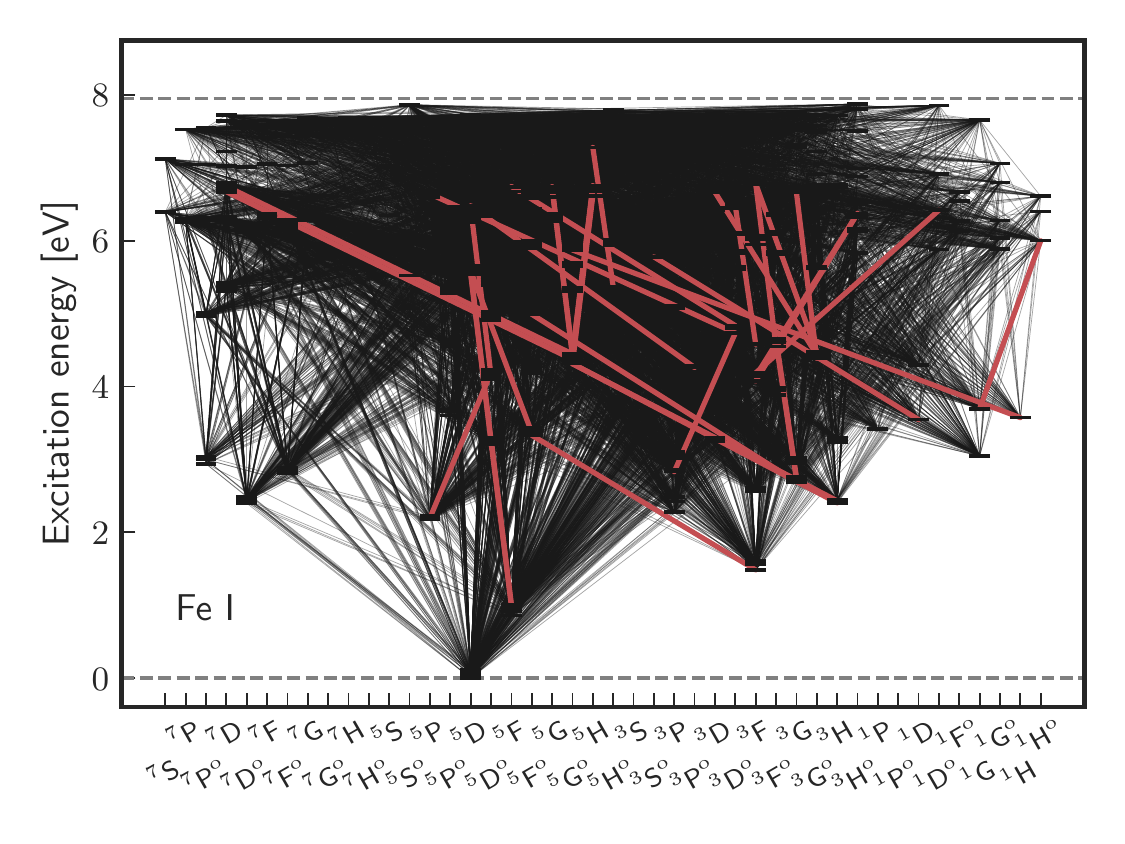}
\includegraphics[width=0.5\linewidth]{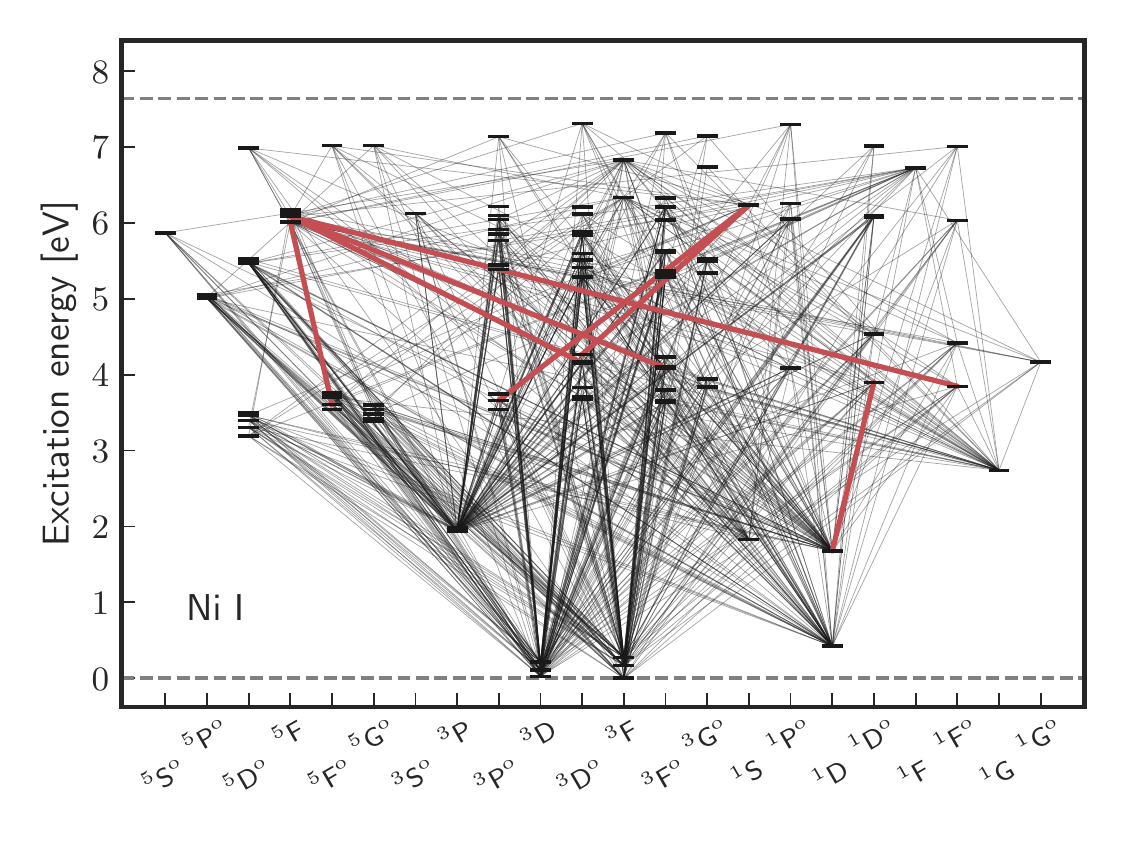}
}
\caption{Grotrian diagrams for chemical elements treated in NLTE. Energy states are depicted with horizontal dashes and connecting bound-bound radiative transitions are shown with black lines. Red lines correspond to diagnostic transitions used in abundance analysis. Only energy levels following L-S coupling are shown.}
\label{fig:grotrian}
\end{figure*} 
\begin{table}
\begin{minipage}{\linewidth}
\renewcommand{\footnoterule}{} 
\setlength{\tabcolsep}{1.5pt}
\caption{Atomic parameters of diagnostic spectral lines of C I, N I, O I, Mg I, Si I, Si II, Ca I, Ni I, and Ni II.}
\label{tab:lines}     
\begin{center}
\begin{tabular}{l c cc ccc}
\noalign{\smallskip}\hline\noalign{\smallskip}  $\lambda$ & $\Elow$ & $\Eup$ & $\log gf$ & vdW\footnote{Van der Waals broadening parameter, see text.} & Ref.\footnote{References: ~~~
(1) \citet{Li2021} 
(2) this work ; 
(3) \citet{Storey2000} 
(4) \citet{PehlivanRhodin2017} 
(5) Henrik Hartman (in preparation, priv. comm.)
(6) \citet{GARZ} renormalized using \citet{BL} 
(7) \citet{SR} 
(8) \citet{S} 
(9) \citet{Wood2014} 
(10) \citet{Johansson2003}}
& Ref.\footnote{References: ~~~(1) \citet{Barklem2000} (2) \citet{Unsold1955} (3) P. Barklem (priv. comm.) (4) \citet{Anstee1991,Anstee1995}} \\
   ~~~~[\AA] & [eV] & [eV] & & & f-val. & vdW \\
\noalign{\smallskip}\hline\noalign{\smallskip}
\ion{C}{i} & & & & & & \\
5052.145 &  7.685 & 10.138 & $-1.36  \pm  0.04$ & $-7.310$ & 1 & 4 \\ 
6587.610 &  8.537 & 10.419 & $-1.05  \pm  0.04$ & 1953.319 & 1 & 1 \\ 
7113.171 &  8.647 & 10.390 & $-0.94  \pm  0.04$ & 1858.314 & 1 & 1 \\ 
\ion{N}{i} & & & & & & \\
8629.235 & 10.690 & 12.126 & $0.006 \pm 0.07$ & 575.234 & 2 & 1 \\ 
8683.403 & 10.330 & 11.757 & $0.162 \pm 0.04$ & 480.231 & 2 & 1 \\ 
\ion{O}{i} & & & & & &  \\
6300.304 & 0.000 &  1.967 & $-9.72   \pm 0.08$  &  $-$ & 3 & 2 \\ 
7771.940 & 9.146 & 10.741 & $0.350  \pm  0.02$ & 453.234 & 2 & 1 \\ 
7774.170 & 9.146 & 10.741 & $0.204  \pm  0.02$ & 453.234 & 2 & 1 \\ 
7775.390 & 9.146 & 10.740 & $-0.019  \pm 0.02$  & 453.234 & 2 & 1 \\ 
\ion{Mg}{i}  & & & & & & \\
5528.405  &  4.346  & 6.588  & $-0.547 \pm 0.02$ & 1461.312  & 4 & 1 \\ 
5711.088  &  4.346  & 6.516  & $-1.742 \pm 0.05$ & 1860.100  & 4 & 3 \\ 
\ion{Si}{i} & & & & & & \\
5645.611 &  4.930 & 7.125 & $ -2.067 \pm 0.03$  & $-7.29$  & 5 & 4 \\ 
5684.484 &  4.954 & 7.134 & $ -1.607 \pm 0.05$  & $-7.30$  & 5 & 4 \\ 
5690.425 &  4.930 & 7.108 & $ -1.802 \pm 0.05$  & 1770.220 & 5 & 1 \\ 
5701.105 &  4.930 & 7.104 & $ -1.981 \pm 0.05$  & 1770.220 & 5 & 1 \\ 
5772.146 &  5.082 & 7.223 & $ -1.643 \pm 0.03$  & $-7.350$ & 5 & 4 \\ 
5793.073 &  4.930 & 7.069 & $ -1.894 \pm 0.1$   & 1700.230 & 5 & 1 \\ 

7034.900 &  5.871 & 7.633 & $-0.78$   &  $-7.13$ & 6 & 4 \\ 
7226.208 &  5.614 & 7.329 & $-1.41$   &  $-7.32$ & 6 & 4 \\ 
\ion{Si}{ii} & & & & & & \\
6371.372 &  8.121 & 10.067 & $-0.120 \pm 0.001$ & $-7.69$ & 5 & 4 \\ 
\ion{Ca}{i} & & & & & & \\
5260.387 & 2.521 & 4.878 & $-1.719 \pm 0.02$ & 421.260  & 7 & 1 \\ 
5512.980 & 2.933 & 5.181 & $-0.464 \pm 0.02$ & $-7.316$ & 8 & 4 \\ 
5867.562 & 2.933 & 5.045 & $-1.570 \pm 0.04$ & $-7.460$ & 8 & 4 \\ 
6166.439 & 2.521 & 4.531 & $-1.142 \pm 0.02$ & 976.257  & 7 & 1 \\ 
6455.598 & 2.523 & 4.443 & $-1.340 \pm 0.04$ & 365.241  & 8 & 1 \\ 
6471.662 & 2.526 & 4.441 & $-0.686 \pm 0.02$ & 365.241  & 7 & 1 \\ 
6499.650 & 2.523 & 4.430 & $-0.818 \pm 0.02$ & 364.239  & 7 & 1 \\ 
\ion{Ni}{i} & & & & & & \\
4740.165 & 3.480 & 6.095 & $-1.72$ & 844.281 &  9 & 1 \\ 
4811.983 & 3.658 & 6.234 & $-1.45$ & $-7.75$ & 10 & 4 \\ 
4814.598 & 3.597 & 6.172 & $-1.63$ & 743.236 &  9 & 1 \\ 
4976.135 & 3.606 & 6.097 & $-1.26$ & 843.282 &  9 & 1 \\ 
5157.980 & 3.606 & 6.009 & $-1.51$ & 691.236 &  9 & 1 \\ 
5537.106 & 3.847 & 6.086 & $-2.22$ & 695.216 &  9 & 1 \\ 
6176.812 & 4.088 & 6.095 & $-0.26$ & 826.284 &  9 & 1 \\ 
6204.604 & 4.088 & 6.086 & $-1.08$ & 719.247 &  9 & 1 \\ 
6223.984 & 4.105 & 6.097 & $-0.91$ & 827.283 &  9 & 1 \\ 
6414.587 & 4.154 & 6.086 & $-1.16$ & 721.249 &  9 & 1 \\ 
\ion{Ni}{ii} & & & & & & \\
6378.250 & 4.154 & 6.097 & $-0.82$ & 825.283 &  9 & 1 \\ 
\noalign{\smallskip}\hline\noalign{\smallskip}
\end{tabular}
\end{center}
\end{minipage}
\end{table}

\begin{table}
\begin{minipage}{\linewidth}
\renewcommand{\footnoterule}{} 
\setlength{\tabcolsep}{1.5pt}
\caption{Atomic parameters of diagnostic Fe lines.}
\label{tab:Fe_lines}     
\begin{center}
\begin{tabular}{l c cc ccc}
\noalign{\smallskip}\hline\noalign{\smallskip}  $\lambda$ & $\Elow$ & $\Eup$ & $\log gf$ & vdW\footnote{Van der Waals broadening parameter, see text.}
& Ref. \footnote{References: ~~~
(1) \citet{BWL}
(2) \citet{MRW}
(3) \citet{2014MNRAS.441.3127R}
(4) \citet{BK}
(5) \citet{WBW} renormalized to \citet{FMW}
(6) average of \citet{BKK}, \citet{GESB82c}, and \citet{BWL}
(7) \citet{GESB82c}
(8) average of \citet{BKK} and \citet{BWL}
(9) average of \citet{GESB82d} and \citet{BWL}
(10) \citet{2009AA...497..611M}
(11) \citet{RU}
}
& Ref. \footnote{References: ~~~ (1) \citet{Barklem2000} (2) \citet{BA-J} } \\
   ~~~~[\AA] & [eV] & [eV] & & & f-val. & vdW \\
\noalign{\smallskip}\hline\noalign{\smallskip}
\ion{Fe}{i} & & & & & & \\
5242.491 & 3.634 & 5.999 & $-0.967 \pm 0.046$ &  361.248 & 1 & 1 \\ 
5365.399 & 3.573 & 5.883 & $-1.020 \pm 0.041$ &  283.261 & 1 & 1 \\ 
5379.574 & 3.695 & 5.999 & $-1.514 \pm 0.046$ &  363.249 & 1 & 1 \\ 
5398.279 & 4.446 & 6.742 & $-0.630 \pm 0.060$ &  993.280 & 2 & 1 \\ 
5543.936 & 4.218 & 6.453 & $-1.040 \pm 0.050$ &  742.238 & 2 & 1 \\ 
5560.212 & 4.435 & 6.664 & $-1.090 \pm 0.050$ &  895.278 & 2 & 1 \\ 
5638.262 & 4.220 & 6.419 & $-0.720 \pm 0.020$ &  730.235 & 3 & 1 \\ 
5661.346 & 4.284 & 6.474 & $-1.765 \pm 0.041$ &  765.209 & 4 & 1 \\ 
5679.023 & 4.652 & 6.835 & $-0.820 \pm 0.050$ & 1106.291 & 2 & 1 \\ 
5731.762 & 4.256 & 6.419 & $-1.200 \pm 0.050$ &  727.232 & 2 & 1 \\ 
5741.848 & 4.256 & 6.415 & $-1.672 \pm 0.097$ &  725.232 & 1 & 1 \\ 
5855.077 & 4.608 & 6.725 & $-1.478 \pm 0.041$ &  962.279 & 4 & 1 \\ 
5905.672 & 4.652 & 6.751 & $-0.690 \pm 0.050$ &  994.282 & 2 & 1 \\ 
5930.180 & 4.652 & 6.742 & $-0.230          $ &  983.281 & 5 & 1 \\ 
6027.051 & 4.076 & 6.132 & $-1.089 \pm 0.051$ &  380.250 & 1 & 1 \\ 
6056.005 & 4.733 & 6.780 & $-0.320 \pm 0.030$ & 1029.286 & 3 & 1 \\ 
6093.644 & 4.608 & 6.642 & $-1.400 \pm 0.080$ &  866.274 & 2 & 1 \\ 
6165.360 & 4.143 & 6.153 & $-1.473 \pm 0.051$ &  380.250 & 1 & 1 \\ 
6187.990 & 3.943 & 5.946 & $-1.620 \pm 0.060$ &  903.244 & 2 & 1 \\ 
6270.225 & 2.858 & 4.835 & $-2.470 \pm 0.059$ &  350.249 & 8 & 1 \\ 
\ion{Fe}{ii} & & & & & & \\
5234.625 & 3.221 & 5.589 & $-2.180$ &  180.249 & 10 & 2 \\ 
5325.553 & 3.221 & 5.549 & $-3.160$ &  179.252 & 10 & 2 \\ 
5425.257 & 3.199 & 5.484 & $-3.220$ &  178.255 & 10 & 2 \\ 
6084.111 & 3.199 & 5.237 & $-3.881$ &  173.223 & 11 & 2 \\ 
6456.383 & 3.903 & 5.823 & $-2.185$ &  185.276 & 11 & 2 \\ 
\noalign{\smallskip}\hline\noalign{\smallskip}
\end{tabular}
\end{center}
\end{minipage}
\end{table}

\subsection{NLTE model atoms} \label{sec:nlteatoms}

New atomic models are available for five chemical elements in our list. We describe them briefly below. The Grotrian diagrams are illustrated in Fig. 
\ref{fig:grotrian}. 

The model atoms of O and Ni are both taken from \citet{Bergemann2021}. In short, the O model includes 122 energy states of O I and O II, coupled by radiative and collision-induced transitions. Radiative data were adopted from the Kurucz\footnote{\url{http://kurucz.harvard.edu/}} database, and supplemented with new photo-ionisation cross-sections for O I states computed using the R-matrix method \citep{Berrington1995}. This method was also used to derive new data for electron-impact transitions. H induced inelastic processes were computed using the OH molecule computed employing the multi-reference configuration interaction (MRCI) method and the collisional dynamics description presented by \citet{Belyaev2019_O_H}. The model atom of Ni primarily relies on the same sources of radiative data (NIST\footnote{\url{https://physics.nist.gov}}, Kurucz), whereas collisional data were calculated using the standard formulae presented in~\citep{Regemorter1962,Seaton1962,Drawin1968}.

The Mg model atom is described in detail in \citet{Zhao1998} and in \citet{Mashonkina2013}, and it was slightly updated in \citet{Bergemann2017}. The model includes 86 energy states, of which 85 represent Mg I, and it is closed by Mg II. Radiative transitions were adopted from the Opacity Project \citep{1995oppr.book.....S}.
453 of these transitions connect the energy levels in Mg I and for 65 states bound-free transition were included. The collisional data were taken from \citet{Mauas1988} and \citet{Zhao1998} for e$^-$ $+$ Mg collisions, and \citet{Barklem2012} was the main source for H I impact excitation and charge exchange processes. Electron-impact ionisation rate coefficients were computed using the \citet{Seaton1962} formula.

The Fe model is described in detail in \citet{Bergemann2012} and it was recently updated by \cite{Semenova2020}. Fe I has probably the most complex system of energy levels of all species in the periodic table. Therefore, representing them in a model is a major numerical challenge.
In the Kurucz database, over 37\,000 energy levels and over 6 million transitions are available.  
The model we employ in this work contains $637$ Fe I states and $58$ Fe II states, which are connected via $19\,267$ radiative  bound-bound transitions. Fine-structure levels of Fe I are included up to $\sim$7 eV. We also include very high-excitation energy levels of Fe I, using the super-levels and super-lines following \cite{Bergemann2012}. The uppermost energy state in Fe I (a super-level) is thus located at 7.88 eV, only 0.07 eV from the first ionisation threshold (7.95 eV) representing the ground state (term) of Fe II. We note that it is common to use fine structure levels for Fe II, however, all ionisation photo-ionisation cross-sections and charge transfer rates are defined for LS states (not for fine structure levels). We tested whether using the actual ionisation thresholds for Fe II levels or representing them as terms has an impact on our results, and found that the difference in abundance space was only 0.01 dex, that is negligible compared to other sources of error in the analysis. We note that the NIST database is very incomplete for the Fe I structure above 7.5 eV.  We therefore also tested more compact atomic models of Fe, devoid of energy levels higher than 7.50 eV in the Fe I system, and we found that using more compact models does not affect the abundances determined from Fe I lines by more than 0.01 dex.

Bound-free radiative cross-sections were taken from \cite{Bautista2017}, whereas the rates of transitions caused by collisions with e$^-$ and H atoms were computed using the new quantum-mechanical estimates of the cross-sections by \cite{Bautista2017} and \cite{Barklem2018}, respectively. The photo-ionisation cross-sections were tabulated in a fine energy grid, to resolve resonances. This is particularly important, because Fe I (and other Fe-group species) react very sensitively to the radiation field, and over-ionisation is indeed the main process behind non-negligible NLTE effects, especially at low metallicity \citep{Bergemann2014, Amarsi2016}.

The atomic model of Si is based on the model presented by \citet{Bergemann2013}, however, with important updates to the radiative and collisional part of the atom. The $f$-values and damping constants for the  Si I lines were substituted by the most recent data available in the Kurucz database\footnote{\url{http://kurucz.harvard.edu/atoms/1400/}, based on Kurucz 2016 calculations}. The rate coefficients describing processes in inelastic collisions between Si I and H atoms were adopted from \citet{Belyaev2014} and from a database by Paul Barklem\footnote{\url{https://github.com/barklem}}. The rate coefficients describing the charge transfer and mutual neutralisation reactions were adopted from \citet{Belyaev2014}.
\begin{table*}[h!]
\begin{minipage}{\linewidth}
\caption{Comparison of $\log gf$-values for the O I transitions. See text.}
\label{table:fvalues}
\centering          
\begin{tabular}{l cc cc c}   
\hline\hline
Reference   & \multicolumn{5}{c}{Transition} \\
      & 777.1 nm & 777.4 nm & 777.5 nm & 630.0 nm &  \Ot{3s}{5}{S}{o}{} - \Ot{3p}{5}{P}{}{} \\
\hline
\citet{Hibbert1991} & & & & & \\
~~CIV3(L)\footnote{CIV3 calculations in the length (L) and velocity (V) gauges} & 0.371 & 0.225 & ~~0.003 & & 0.702 \\
~~$\mathrm{CIV3(V)}^{a}$ & 0.333 & 0.187 & $-0.035$ & & 0.664 \\
& & & & & \\
\citet{Joensson2000} & & & & & \\
~~MCHF(L)\footnote{Non-relativistic multiconfiguration Hartree-Fock (MCHF) calculations in the length (L) and velocity (V) gauges} & & & & & 0.682 \\
~~$ \mathrm{MCHF(V)}^{b}$ & & & & & 0.679 \\
 & & & & & \\
\citet{Civis2018} & & & & & \\
~~QDT\footnote{Quantum Defect Theory (QDT)} & 0.317 & 0.170 & $-0.051$ & & 0.647 \\
 & & & & & \\
\citet{Storey2000} & & & & & \\
~~SST\footnote{SUPERSTRUCTURE (SST) calculations}  & & & & $-9.72$ & \\
 & & & & & \\
This work & & & & & \\
~~HFR$+$CPOL\footnote{Pseudo-relativistic Hartree-Fock method including core-polarization effects (HFR$+$CPOL)}     & 0.350 & 0.204 & $-0.018$   & $-9.65$ & 0.681 \\
 & & & & & \\
~~MCDHF(B)\footnote{Fully relativistic multiconfiguration Dirac-Hartree-Fock (MCDHF) calculations in the Babushkin (B) and Coulomb (C) gauges}  & 0.370 & 0.224 & ~~0.002   & & 0.701 \\
~~$\mathrm{MCDHF(C)}^{f}$       & 0.331 & 0.184 & $-0.039$ & & 0.662 \\
~~$\mathrm{MCDHF}^{f}$ & & & & $-9.69$ & \\
 & & & & & \\

~~AST(L)\footnote{AUTOSTRUCTURE (AST) calculations in the length (L) and velocity (V) gauges}  & 0.348 & 0.199 & $-0.021$ & \\
~~$\mathrm{AST(V)}^{g}$  & 0.326 & 0.176 & $-0.043$ & \\
 & & & & & \\

\hline
Final recommended & & & & & \\
     & 0.350 & 0.204 & $-0.019$ & $-9.69$ & \\
\hline                  
\end{tabular}
\end{minipage}
\end{table*}
\subsection{New oscillator strengths for N and O}\label{sec:newfvalues}

To compute the new values of oscillator strengths, we adopt different independent approaches.

For the O I lines, we used the pseudo-relativistic Hartree-Fock method, originally introduced by \citet{Cowan1981} modified to account for core-polarization effects (HFR+CPOL), as described e.g. by \citet{Quinet1999} and \citet{Quinet2002} and the fully relativistic multiconfiguration Dirac-Hartree-Fock (MCDHF) approach developed by \citet{Grant2007} and \citet{Froese2016} with the latest version of GRASP (General Relativistic Atomic Structure Program), known as GRASP2018 \citep{Froese2019}. In our HFR+CPOL calculations, the valence-valence interactions were considered among a set of configurations including 2p$^4$, 2p$^3$ $nl$ (with $nl$ up to 6h) and all single and double excitations from 2s, 2p to 3s, 3p, 3d orbitals while the core-valence correlations were modelled by a core-polarization potential corresponding to a He-like O VII ionic core with a dipole polarizability $\alpha_d$ = 0.0026 a$_0^3$ \citep{Johnson1983} and a cut-off radius $r_c$ = 0.198 a$_0$. The latter value corresponds to the calculated HFR value of $<$$r$$>$ of the outermost core orbital (1s). For the MCDHF calculations, we adopted a physical model in which valence-valence correlations were considered by means of single and double excitations from the 2p$^4$, 2p$^3$3s, 2p$^3$3p, 2p$^3$3d multireference (MR) configurations to the \{9s,9p,9d,6f,6g,6h\} orbital active set (where the maximum orbital principal quantum number $n_{max}$ is specified for each orbital azimuthal quantum number $l= s-h$) while the core-valence effects were included by means of single and double excitations from 1s to the active set of orbitals. The convergence of the oscillator strengths for the O I lines was verified by comparing the results obtained using physical models including increasingly large active sets of orbitals, and by observing a good agreement  between the $gf$-values computed within the Babushkin (B) and the Coulomb (C) gauges. The two gauges tend to preferentially weight different parts of the wavefunctions (that is the outer region vs. near-nucleus), and so can be used to quantify the systematic effects. 

We also performed large, exhaustive calculations of $f$-values for the diagnostic lines of O I and N I using the code AUTOSTRUCTURE \citep{2011AIPC.1344..139B}. We considered all one-, two-, and three-electron promotions from the 2s and 2p orbitals of configurations 2s$^2$2p$^4$ to excited orbitals $nl$, with $3\le n \le 6$ and $0\le l\le 4$. We analyzed the convergence or lack thereof of the oscillator strengths with increasingly large configuration expansions, which accounted for hundreds of configurations for each element. In every calculation we made use of term energy corrections and level energy corrections, which are semi-empirical corrections available in AUTOSTRUCTURE. Convergence of calculated oscillator strength values was evaluated by two criteria that are the stability of the numeric values as more configurations are accounted for in the configuration interaction expansion and the agreement between calculated length and velocity gauges of the oscillator strength.

\renewcommand{\footnoterule}{} 
\begin{table*}[h!]
\begin{minipage}{\linewidth}
\begin{center}
\caption{Solar photospheric abundances derived in NLTE. See the text for details.}
\label{table:1d3d_abundances}
\centering          
\begin{tabular}{lcccc}   
\hline\hline      
   El. & A(El), $\sigma_{\rm tot}$ & A(El), $\sigma_{\rm tot}$ & $\Delta_{\rm STAGGER - CO5BOLD}$  & $\sigma_{\rm stat}$ \\
           & 1D\footnote{1D MARCS model atmosphere} NLTE   & $\td$\footnote{$\td$ \stagger\ model atmosphere} NLTE \footnote{LTE for C and N}     & \\
\hline 
   C &  $ 8.48 \pm 0.08$ &   $ 8.56 \pm 0.06  $  & ~~$0.016$ & $ 0.05 $ \\ 
   N &  $ 7.88 \pm 0.12$ &   $ 7.98 \pm 0.10  $  & $-0.011$  & $ 0.1 $ \\
   O &  $ 8.74 \pm 0.04$ &   $ 8.77 \pm 0.04  $  & ~~$0.014$ & $ 0.02 $ \\
   Mg & $ 7.45 \pm 0.08$ &   $ 7.55 \pm 0.06  $  & ~~$0.020$ & $ 0.06 $ \\
   Si & $ 7.54 \pm 0.07$ &   $ 7.59 \pm 0.07  $  & ~~$0.005$ & $ 0.06 $ \\
   Ca & $ 6.34 \pm 0.05$ &   $ 6.37 \pm 0.05  $  & ~~$0.026$ & $ 0.04 $ \\
   Fe & $ 7.49 \pm 0.08$ &   $ 7.51 \pm 0.06  $  & ~~$0.012$ & $ 0.08 $ \\
   Ni & $ 6.21 \pm 0.04$ &   $ 6.24 \pm 0.04  $  & ~~$0.001$ & $ 0.07 $ \\
\hline                  
\end{tabular}
\end{center}
\end{minipage}
\end{table*}

The comparison of the f-values with the literature estimates is provided in Tab. \ref{table:fvalues}. The values obtained with the MCDHF method and with AUTOSTRUCTURE agree to better than 1.5\,$\%$ for the velocity gauges, and 5 $\%$ for the length gauges. Our results are also in a good agreement with the literature, including the values published by \citet{Hibbert1991} and by \citet{Joensson2000}. 

Our recommended values for the O I triplet lines are based on our new estimates calculated with both the MCDHF approach and AUTOSTRUCTURE. The final values of oscillator strengths are $\log gf = 0.350, 0.204$, and $-0.019$ dex for the 777.1, 777.4, and 777.5 nm transitions, respectively. These f-values are nearly identical to those adopted by \citet{Bergemann2021}, with a difference of $1.6\%$ only.
The 630 nm line is an M1 transition and the transition operator is gauge independent. For this transition, our best value is the MCDHF result ($\log gf = -9.69$ dex), because it includes relativistic and electron correlation effects in more detail than the SST and HFR$+$CPOL calculations.
The differences between different methods and between the two gauges are used to estimate the uncertainties of our $f$-values using the standard expression used in atomic physics $dT = |gf(\rm{B})-gf(\rm{C})|/max(gf(\rm{B}),gf(\rm{C}))$. It is important to note that there is no reason to prefer one gauge over the other. The convergences of the f-values in both of them demonstrate a similar behaviour. We therefore advocate to use the averages of results obtained in length and velocity form, or their relativistic equivalents.

\subsection{Line lists}\label{sec:linelists}

We make use of the line list compiled for the Gaia-ESO survey (GES) described in \citet{Heiter2021}, which includes basic atomic data (wavelengths, energy levels, transition probabilities, broadening constants) and quality flags for lines relevant to FGK stars in the wavelength range from $475$ to $685$ nm and from $849$ to $895$ nm. The line list was assembled by a dedicated GES working group after a detailed assessment of all available sources of atomic and molecular data, with priority given to experimental data and carefully validated theoretical data. For a subset of atomic transitions tabulated in the line list a pair of flags is given for each transition that indicates the quality of the atomic data (primarily, $\log gf$) and the quality of the line in the stellar spectrum (primarily, blending). For each of these two aspects, the lines were sorted into three categories of decreasing quality, designated 'Y', 'U', and 'N'. The quality assessment was based on a comparison of synthetic and observed spectra for the Sun and Arcturus.

Since the GES line list only includes atomic data published until 2014, we updated the entries whenever more recent datasets have become available in the literature. The data and references for all the spectral lines used for the abundance calculations are given in Tables~\ref{tab:lines} and \ref{tab:Fe_lines}. For the diagnostic lines, van der Waals broadening parameters were adopted, where available, from \citet{Barklem2000}. Also for the other lines in the linelist, we gave preference to the Anstee-Barklem-O’Mara theory
(\citealt{Anstee1991} 
and successive expansions by P.S.\ Barklem and collaborators). Tables~\ref{tab:lines} and \ref{tab:Fe_lines} give the values in a packed notation where the integer component is the broadening cross-section, $\sigma$, in atomic units, and the decimal component is the dimensionless velocity parameter, $\alpha$. Values less than zero are the logarithm of the broadening width per unit perturber number density at 10\,000~K in units of rad s$^{-1}$ cm$^{3}$.

The oscillator strengths were adopted from the following sources. For \ion{C}{i}, we used the new $f$-values from \citet{Li2021}, which were computed by means of multiconfiguration Dirac-Hartree-Fock and relativistic configuration interaction methods. We note that for most transitions they are systematically lower compared to the commonly used values from \citet{Hibbert1993}. For the \ion{Mg}{i} lines at 5528~\AA\ and 5711~\AA\ we used the experimental $gf$-values from \citet{PehlivanRhodin2017}. For the \ion{Ca}{i} line at 6455.598~\AA\ we updated the value in the GES line list to that recommended in \citet{DenHartog2021}. Also the other \ion{Ca}{i} lines that we used are included in the list of recommended lines by \citet{DenHartog2021}, and the uncertainties quoted in Table~\ref{tab:lines} are those recommended by \citet{DenHartog2021}\footnote{These uncertainties refer to the absolute $gf$-values, while the uncertainties given in the GES line list are those of the relative measurements quoted in \citet{SR} and \citet{S}.}.
Most of the lines used for the abundance calculations that are also included in the GES line list have $gf$-value quality flag 'Y', except for a few Fe lines with flag 'U'. Concerning the blending quality flag, most lines have flag 'Y' or 'U', except for two \ion{Ni}{i} lines (at 4811.983 and 4814.598~\AA) with flag 'N'. However, these two lines are clearly blended only in the spectrum of Arcturus, and much less so in the spectrum of the Sun.

For the wavelength ranges not covered by the GES line list, we used data from the VALD database\footnote{\url{http://vald.astro.uu.se}} \citep{Pisk:95,2015PhyS...90e4005R}, 
and complemented them with molecular line lists provided by B. Plez (priv. comm.). It should be noted, though, that special care was taken to select only lines minimally affected by blends. Therefore the inclusion of molecules in the spectrum synthesis is strictly speaking not necessary, except for the modelling of \ion{N}{i} lines, which are blended by CN features.
\section{Results}\label{sec:abundances}
Our results for all relevant chemical elements are collected in Table~\ref{table:1d3d_abundances}, and in Table~\ref{table:final-abundances} we provide the abundances of all species that are relevant to the SSM. To illustrate the quality of the spectral fits, we show selected examples of the best-fit spectra compared to the observed profiles in Fig. \ref{fig:linefit2}. Abundances measured from individual lines are presented in Table~\ref{tab:ind_abundances} in Appendix~\ref{sec:appendix}.
\begin{table}[h!]
\begin{minipage}{\columnwidth}
\renewcommand{\footnoterule}{} 
\setlength{\tabcolsep}{1.5pt}
\caption{Our recommended solar chemical abundances on the astronomical scale $A(i) = \log{(N(i)/N_H)}+12$. Uncertainties listed for the meteoritic abundances are purely statistical but a systematic and fully correlated 0.02~dex uncertainty should be included for all elements (Sect.~\ref{sec:meteo}). The last row reports the total Z/X, where the abundances of volatile elements are always taken from the photospheric scale. For the meteoritic scale, the second uncertainty source reflects the fully correlated uncertainty from transforming the cosmochemical scale to the astronomical scale. See the text.}
\label{table:final-abundances}
\begin{center}          
\begin{tabular}{l cc c}   
\hline\hline      
   Element & A(El)$_{\rm ph}$, $\sigma_{\rm tot}$ & Ref.\footnote{References: ~~~
(1) this work 
(2) \citet{Maiorca2014}
(3) \citet{Zhao2016}
(4) \citet{Nordlander2017}
(5) \citet{Scott2015a}
(6) \citet{Caffau2011}
(7) \citet{Lodders2019}
(8) \citet{Bergemann2011}
(9) \citet{Scott2015b}
(10) \citet{Bergemann2010b}
(11) \citet{Bergemann2019}
(12) \citet{Bergemann2010a}
}
& A(El)$_{\rm met}$, $\sigma_{\rm stat}$ \\
\hline 
   H  & 12.0             & - & - \\
   C  & $8.56 \pm 0.05$  & 1 & - \\
   N  & $7.98 \pm 0.08$  & 1 & - \\
   O  & $8.77 \pm 0.04$  & 1 & - \\
   F  & $4.40 \pm 0.25$  & 2 & $4.67 \pm 0.09$ \\
   Ne & $8.15 \pm 0.11$  & 1 & - \\
   Na & $6.29 \pm 0.02$  & 3 & $6.33 \pm 0.04$ \\
   Mg & $7.55 \pm 0.05$  & 1 & $7.58 \pm 0.02 $ \\
   Al & $6.43 \pm 0.03$  & 4 & $6.48 \pm 0.03 $ \\
   Si & $7.59 \pm 0.07$  & 1 & $7.57 \pm 0.02 $ \\   
   P  & $5.41 \pm 0.03$  & 5 & $5.48 \pm 0.03 $ \\
   S  & $7.16 \pm 0.11$  & 6 & $7.21 \pm 0.03 $ \\
   Cl & [$5.25 \pm 0.12$]& 7 & $5.29 \pm 0.07 $ \\
   Ar & [$6.50 \pm 0.10$]& 7 & - \\
   K  & $5.14 \pm 0.10$  & 3 & $5.12 \pm 0.02 $ \\
   Ca & $6.37 \pm 0.04$  & 1 & $6.32 \pm 0.03 $ \\
   Sc & $3.07 \pm 0.04$  & 3 & $3.09 \pm 0.03 $ \\
   Ti & $4.94 \pm 0.05$  & 8  & $4.96 \pm 0.03 $ \\
    V & $3.89 \pm 0.08$  & 9  & $4.01 \pm 0.03 $ \\
   Cr & $5.74 \pm 0.05$  & 10 & $5.69 \pm 0.02 $ \\
   Mn & $5.52 \pm 0.04$  & 11 & $5.53 \pm 0.03 $ \\
   Fe & $7.50 \pm 0.06$  & 1  & $7.51 \pm 0.02 $ \\
   Co & $4.95 \pm 0.04$  & 12 & $4.92 \pm 0.02 $ \\
   Ni & $6.24 \pm 0.04$  & 1  & $6.25 \pm 0.03 $ \\
\hline                  
Z/X           & $0.0225 \pm 0.0014$ & & $0.0226 \pm 0.0014 \pm 0.0003$ \\
\noalign{\smallskip}\hline\noalign{\smallskip}
\end{tabular}
\end{center}
\end{minipage}
\end{table}

\begin{figure*}[h!]
\hbox{
\includegraphics[width=0.99\linewidth]{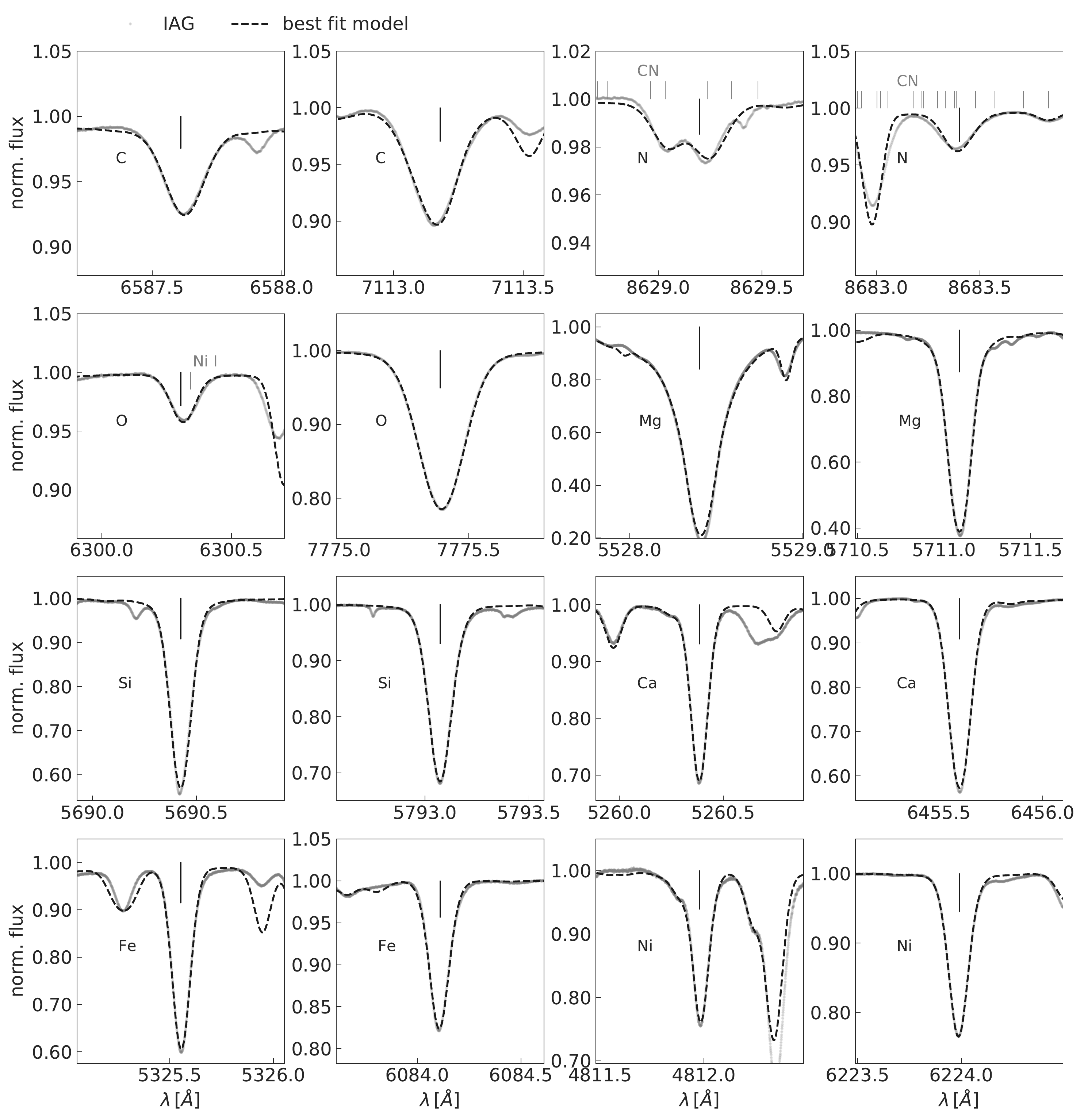}
}
\caption{Comparison of model and observed line profiles for some of the diagnostic lines.}
\label{fig:linefit2}
\end{figure*} 
The uncertainties of the abundances were derived in analogy to \cite{Bergemann2021}. We form uniform distributions of errors using the values representing a) the uncertainty of the transition probability ($f$-value), b) the systematic uncertainty caused by using different $\td$ models, c) the uncertainty caused by the H collisional data (only for elements treated in NLTE), and d) the statistical uncertainty. The latter is represented by the scatter of abundances derived from individual spectral lines of the same chemical element and is stated in the last column of Table~\ref{table:1d3d_abundances}. The uncertainty caused by using different $\td$ models is represented by the difference between the abundances derived with $\td$ STAGGER and CO5BOLD models, see fourth column of Table~\ref{table:1d3d_abundances}. The error distributions are then co-added, as the uncertainties are independent. Since the shapes of the distributions are close to Gaussian, we adopt the simple averages and the standard deviations ($1\sigma$) of the resulting combined distributions as our final abundances and their corresponding uncertainties, respectively.
\subsection{New estimates of photospheric abundances}
\subsubsection{C}
Our analysis of the C abundance relies on atomic C lines, because these are the features we expect to be able to measure in the spectra of upcoming facilities, such as 4MOST and WEAVE. Since the diagnostic lines of C I are almost insensitive to NLTE (see below), we rely on $\td$ LTE calculations to obtain the solar C abundance of A(C) $= 8.56 \pm 0.06$ dex. For comparison, the values obtained by  \citet{Amarsi2019_C} and \citet{Caffau2010} are $8.44 \pm 0.02$ dex and $8.50 \pm 0.06$ dex, respectively, the latter in agreement with our work. Also, the revised 3D NLTE abundance by \cite{Li2021} is $8.50 \pm 0.07$ dex, which is consistent with our value. We note that using the new f-values calculated by \citet{Li2021}, we obtain a significantly improved agreement between different C I lines, compared to the result obtained using the older f-values from \citet{Hibbert1993}. The line-by-line scatter drops from $0.08$ dex (with \citealt{Hibbert1993} values) to $0.05$ dex (with \citealt{Li2021} values). Especially, the much lower $\log gf$ for the 7113.18 \AA~line found by the latter study is essential to bring line into agreement with the other optical lines of C I. This result suggests that the new f-values from \citet{Li2021} are more reliable than the older values from \citealt{Hibbert1993}.
 
Another estimate of the solar C abundance was recently presented by \citet{Alexeeva2015}, who obtained (in 3D) A(C) $= 8.43 \pm 0.02$ dex from the analysis of the CH lines, A(C) $= 8.46 \pm 0.02$ dex from the C$_2$ lines, A(C) $= 8.43 \pm 0.03$ dex from the C I permitted lines, and A(C) $= 8.45$ dex from the forbidden [C~I] feature. However, these estimates rely on older \citet{Hibbert1993} data, and after re-normalisation to the \citet{Li2021} f-values\footnote{The values from \citet{Li2021} are typically $0.05$ dex to $0.16$ dex lower for the diagnostic optical lines of C I. Hence, the C abundance is correspondingly higher.} their abundance is in excellent agreement with our result. Our analysis of C abundances is carried out in LTE, however, according to 3D NLTE calculations by \citet[][their Table~2]{Amarsi2019_C}, 3D NLTE and 3D LTE abundances based on our diagnostic lines of C I agree within $0.01$ dex.

Our estimate of the C abundance obtained using the CO5BOLD model is less than $0.02$ dex lower compared to that calculated using the STAGGER model. This suggests that the difference between \citet{Caffau2010} and \citet{Amarsi2019_C} is likely not caused by the differences between the two 3D solar model atmospheres. The value provided by \citet{Asplund2021} is A(C) $= 8.46 \pm 0.04$ dex, somewhat higher compared to the previous estimate by \citet{Asplund2009}.
\subsubsection{N}
Our estimate of the N abundance relies on the modelling of the two least-blended N I lines in the solar spectrum. For the lack of an NLTE model of N, we resort to the LTE analysis. The recent study by \citet{Amarsi2020} suggests that the difference between 1D LTE and $\td$ NLTE abundances for the optical N I lines are extremely small, and does not exceed $0.005$ dex. In this work, we employ the new $f$-values calculated as described in Sect.~\ref{sec:newfvalues}. Our recommended value of the solar N abundance thus becomes A(N) $= 7.98 \pm 0.10$ dex.
The study by \cite{Amarsi2020} advocated for the solar N abundance of A(N) = $ 7.77 \pm 0.05$ dex, while \cite{Asplund2021} derived N abundance based on both atomic and molecular lines, A(N) = $7.83 \pm 0.07$, with a value unchanged from that of \cite{Asplund2009}, where A(N) = $ 7.83 \pm 0.05$ dex. We note, however, that the N abundance has a very minor impact on the overall metallicity, see Sect.~\ref{sec:ssm} and Fig. \ref{fig:dcdz}.

In the choice of the solar N value, it is important to stress that in this work we rely on ab-initio atomic and molecular data, and we do not apply any empirical adjustment to the line lists. In particular, all molecular features are included self-consistently in the radiative transfer and spectrum synthesis calculations. In contrast, the solar N abundance provided by \citet{Amarsi2020} is based on re-scaling the strengths of the CN blends in a semi-empirical approach, by estimating their equivalent widths in the solar disc-center spectrum through a comparison with nearby CN lines. Whereas both approaches have their pro's and contra's, the main value of our strictly theoretical approach is that it is universal and can be applied to any star, not biasing the result by the assumed value of the observed disc-center equivalent widths of the CN blends. 
We find that abundance obtained using the CO5BOLD model is only $0.011$ dex higher compared to that calculated using the STAGGER model.
\subsubsection{O}
Our result for O, A(O) $=8.77 \pm 0.04$ dex, is in excellent agreement with the full 3D NLTE analysis by \citet[][8.75 $\pm$ 0.03 dex]{Bergemann2021}. The difference between the two results, which is not significant, is caused by using the flux spectrum and the average $\td$ atmosphere model in this work.

Our O abundance is also consistent with the value by \citet[][8.76 $\pm$ 0.07 dex]{Caffau2008}, but it is somewhat higher compared to the estimate by \citet[][8.69 $\pm$ 0.04 dex]{Asplund2021} and \citet[][8.70 $\pm$ 0.04 dex]{Amarsi2021}. These differences can be attributed to the use of new $\log gf$ values for the O lines, new observational material, and our first self-consistent NLTE radiative transfer for both O and Ni features, which is important for the critical diagnostic [O~I] feature at 630~nm. 
As investigated in \citet{Bergemann2021} in detail, quantitatively, the break-down of systematic difference with 
\citet{Asplund2021} is as follows. Our new f-values lead to $+0.02$ dex higher abundances for the 777 nm lines, but $-0.03$ dex lower values for the 630 nm [O I] line. The IAG data further lead to a $+\sim 0.02$ higher abundance for the 777 nm and $+0.03$ dex higher values compared to the SST data, the SST data are however affected by the problem of fringing (Sect.~4.4 in \citet{Bergemann2021}). Modelling Ni in NLTE leads to a $\sim 0.03$ dex higher O abundance inferred from the [O $+$ Ni] feature at 630 nm. There are also very minor differences caused by the continuum placement, model atmospheres, and resolution in radiative transfer modelling. Our results (Table~\ref{tab:ind_abundances} for the 630 nm and 777 nm lines are in excellent agreement.

Overall, our new O abundance is closer to the classical values from \citet{Grevesse1998} than the previous estimate by \citet{Asplund2021}. We also note that according to \citet{Caffau2013}, there is still an unresolved mismatch between the 630~nm and 636~nm [O~I] lines, with the latter line yielding an the O abundance of $8.78 \pm 0.02$ dex. 

To estimate the difference between the STAGGER and CO5BOLD models, we have furthermore performed a comparative analysis of the results obtained with both simulations. We find that the CO5BOLD model leads to a $0.006$ dex lower abundance for the 630~nm forbidden line, and to $0.015$ dex lower abundance for the permitted 777~nm O I triplet. This difference is not significant enough to explain the differences between the \citet{Caffau2008} and \citet{Asplund2009} results.
\subsubsection{Mg}
Our estimate of the solar photospheric abundance of Mg is A(Mg) = $7.55 \pm 0.06$ dex. Comparing with the value obtained by \citet{Bergemann2017}, $7.56 \pm 0.05$ dex, we find a very good agreement. Our results also support the recent detailed analysis of Mg in NLTE by \citet{Alexeeva2018}, who find A(Mg)$=7.54 \pm 0.11$ dex based on Mg I and A(Mg)$=7.59 \pm 0.05$ dex based on Mg II lines. The comparison of abundances obtained from the CO5BOLD and STAGGER $\td$ models (Table~\ref{table:1d3d_abundances}) suggests that the atomic Mg lines used in this work (at 5528 and 5711 \AA) are not sensitive to the differences between the $\td$ model atmospheres, with the difference in abundance not exceeding $0.02$ dex.

The main uncertainty in the analysis of Mg abundances is still associated with the errors of the oscillator strengths and damping parameters. For comparison, the $\td$ NLTE solar Mg abundance from \citet{Osorio2015} is considerably higher, $7.66 \pm 0.07$ dex. The offset could possibly be explained by the differences in the model atom, adopted atomic data, and the choice of lines in the calculations. 
\citet{Osorio2015} find significant ($+0.05$ to $+0.15$ dex) differences between the Mg abundances derived from the Mg I lines using the 1D and $\td$ solar model atmospheres. This is confirmed by our analysis.
\subsubsection{Si}
The Si abundance is a very important parameter in the analysis, as the element has traditionally been used as the main anchor between the solar photospheric and the meteoritic abundance scales \citep{Lodders2003,Asplund2009,Lodders2019}. Our $\td$ NLTE value, A(Si) $ = 7.59 \pm 0.07$ dex is slightly higher compared to the recent 3D NLTE estimate by \citet{Amarsi2017} and more recently by \citet{Asplund2021}. 

For the Si I lines, the NLTE effects are negative, such that NLTE abundances are lower compared to LTE values \citep{Shi2008,Bergemann2013,Mashonkina2020}, which is also supported by our new results. Our NLTE abundances for all lines in the list are $\sim 0.03$ to $0.05$ dex lower compared to LTE. The most recent NLTE estimate of the solar Si abundance, based on the NLTE line-by-line spectrum synthesis, was presented in \citet{Mashonkina2020}. They find that NLTE abundances obtained from the solar Si I lines are on average $0.02$ to $0.1$ dex lower compared to LTE (their Table~2), fully in agreement with our findings.  Using their results for the lines in common with our study and re-normalising them to employed in this study $f$-values, their A(Si) becomes $7.55$ dex, fully in agreement with our 1D NLTE estimate. In fact, their solar Si abundance using their preferred $f$-values would be $7.60$ dex for the lines in common between our and their study, thus about $0.1$ dex higher compared to  \citet{Asplund2021}.

Si I lines are similar to Mg I in that the lines are barely affected by the detailed structure of the 3D model atmosphere. The difference between the results based on CO5BOLD and STAGGER does not exceed $0.005$ dex.
\subsubsection{Ca}
The solar photospheric abundance of Ca is robust, as our $\td$ NLTE calculations show an excellent consistency between different diagnostic lines of Ca I. Our recommended value is A(Ca) $=6.37 \pm 0.05$ dex. This value is slightly higher compared to the recommended value from \citet[][$6.30 \pm 0.03$ dex]{Asplund2021}. 

Other recent estimates of the solar Ca abundances were presented by \citet{Mashonkina2017} and \citet{Osorio2019}. \citet{Mashonkina2017}, performed a 1D NLTE analysis with the solar MARCS model, which resulted in solar abundances of A(Ca) $= 6.33 \pm 0.06$ dex based on Ca I lines and A(Ca) $= 6.40 \pm 0.05$ dex based on Ca II lines with high values of $\Elow$. The average of these quantities is consistent with our $\td$ NLTE estimate, which is not surprising as their NLTE model atom and the chosen line list are similar to our inputs.

\citet{Osorio2019} also used 1D hydrostatic models in combination with NLTE line formation. However, their solar abundances show a significant scatter, ranging from A(Ca) = $6.0$ dex to $6.45$ dex, which likely results from the choice of lines in their analysis.
\subsubsection{Fe}
Our results for the solar Fe abundance are close the recent literature estimates. In $\td$ NLTE, we obtain A(Fe) $= 7.51 \pm 0.06$ dex, which represents the combination of the abundance determined from the Fe I and Fe II lines. The abundances derived from the lines of the neutral species are sensitive to NLTE, which is also known from literature studies \citep[e.g.][]{Bergemann2012, Lind2012}. However, the majority of diagnostic lines in our line list have $\Elow \gtrsim 2.5$ eV, therefore the difference with the LTE abundance is not large and does not exceed $0.1$ dex for the majority of Fe I lines. 
We do not include Fe I lines with low $\Elow$ values in the abundance analysis, due to their high sensitivity to the choice of model atmosphere, see Sect.~\ref{sec:stagger_cobold_comp}. However, this choice of Fe lines affects the final Fe abundance by no more than $0.01$ dex. 
Likewise, the abundance differences obtained from the CO5BOLD and STAGGER models do not exceed $0.012$ dex.

Comparing our result with the recent estimates by two different groups, \citet[][$7.48 \pm 0.04$ dex]{Lind2017} and \citet[][$7.54$ dex]{Mashonkina2019}, we find a good agreement with both studies. The latter estimate is based on the analysis of Fe II lines using the solar MARCS model atmosphere and the $f$-values from \citet{Raassen1998}, whereas the former relies on 3D NLTE modelling of the lines of both ionisation species. Also the careful systematic analysis of Fe lines in NLTE by \citet{Sitnova2015} supports our result, although they relied on the Drawin formula to describe Fe$+$H collisions instead of quantum-mechanical data. 

Our solar Fe abundance is also fully consistent with the earlier estimate by \citet{Caffau2007}, who found A(Fe)$ = 7.52 \pm 0.06$ dex, they also suggest that the result is not sensitive to the choice of $f$-values. The value from \citet[][$7.46 \pm 0.04$ dex]{Asplund2021} is, in contrast, significantly lower compared to our values, to \citet{Caffau2007}, and to \citet{Mashonkina2019}. 
\subsubsection{Ni}
This work presents the first detailed analysis of the solar photospheric abundance of Ni using $\td$ NLTE calculations. Our best estimate is A(Ni) = $6.24 \pm 0.04$ dex, and it is based on the detailed modelling of 11 lines of Ni I and one line of Ni II using the $f$-values from \citet{Wood2014}. 

We do not detect any systematic trend with the excitation potential of the lower energy level or other atomic parameters, which provides confidence in the results. The NLTE effects in the diagnostic solar Ni lines are modest: for the Sun, the differences between 1D NLTE and 1D LTE results are of the order $0.01$ dex. However, as shown in \citet{Scott2015b} and in \citet{Bergemann2021}, the lines of Ni are sensitive to the temperature structure of the model, and the calculations with models based on 3D RHD simulations yield somewhat higher Ni abundance in LTE and NLTE. 

We also find that Ni lines are not very sensitive to the structure of the $\td$ model atmospheres. The abundances derived with \stagger\, model are within 0.02 dex different from the abundances derived using temporarily spaced \cobold\, snapshots. However, these differences average out when combining results from  individual \cobold\ snapshots. So the final difference between \stagger\, and \cobold\ is negligible, below $0.001$ dex.
\subsection{Other chemical elements}
\subsubsection{Ne, Cl, Ar}
The analysis of the Ne abundance is beyond the scope of this paper, because the abundance of the element cannot be determined from the solar photospheric spectrum. It is common \citep[e.g.][]{Lodders2019} to determine the solar Ne abundance from the Ne/O ratio that can be established independently and by assuming the ratio is the same in the solar photosphere, the absolute Ne abundance is derived. 

This procedure was adopted, e.g, by \citet{Lodders2003} to derive the Ne abundance by combining the Ne/O ratio measured in young B type stars and He II regions \citep[][]{Meyer1989}, in solar active regions  \citep[][]{Widing1997}, and in solar energetic particle events \citep[][]{Reames1998} with the photospheric O abundances. Using the Ne/O ratio from \citet{Lodders2003} (Ne/O $= 0.152$), and adopting our O abundance, we would obtain A(Ne) $= 7.95 \pm 0.1$ dex. This value is consistent with that by \citet{Bochsler2007}, who obtained the Ne abundance by combining the solar wind data from the APOLLO foil experiment \citep{Geiss1972} with a model for the Coulomb-drag fractionation in the solar wind. \citet{Juett2006} determined the Ne/O ratio in the interstellar medium (ISM, $0.185$) from the K- and L-shell spectroscopy of nine X-ray binaries. Whether the solar values and the ISM values are consistent is still debated, as the latter is, in particular, prone to systematic effects, such as the choice of the model describing the ionization structure of the Local Bubble \citep[e.g.][]{Breitschwerdt2021}. 
The recent analysis of the local ISM data acquired with the IBEX satellite by \citet{Park2014} indicates a significantly higher Ne/O value of $0.33$. This estimate supports the earlier values based on ISM data collected with the SWICS spectrometer on Ulysses (Ne/O $= 0.26$ dex, \citealt{Gloeckler2007}, their Table~3, estimated based on total densities of Ne and O in the local interstellar cloud). It is also consistent with the estimate based on the analysis of \textit{Chandra} X-ray spectra (Ne X, O VIII, OVII) of nearby FGK-type main-sequence stars by \citet[][0.41]{Drake2005}. Furthermore, a recent NLTE analysis of $24$ B-type stars by \citet{Alexeeva2020} finds non-negligible differences with Ne abundances presented by \citet[][]{Meyer1989}.

In the most recent analysis of the SOHO data with new atomic data, \citet{Young2018} found a Ne/O ratio of $0.24 \pm 0.05$ dex. Using their value, we obtain A(Ne)$=8.15$ dex, which is higher than the estimate calculated using the Ne/O fraction from \citet{Lodders2003}. In the light of remaining uncertainties associated with the ISM and B-type diagnostics, we have opted for determining our final Ne abundance using the \citet{Young2018} Ne/O ratio, which is based on the SOHO measurements of emission lines from the transition region of the quiet Sun. That value is about 40 $\%$ higher compared to an earlier estimate by the same author \citep{Young2005}, because of the use of more accurate ionization and recombination rate coefficients. We adopt a more conservative error to reflect the uncertainty associated with the latter, as well as the uncertainty of our photospheric O abundance. Our recommended abundance is, thus, A(Ne) $=8.15 \pm 0.11$ dex. The value is consistent with the estimate by \citet[][]{Young2018}, A(Ne) $= 8.08 \pm 0.09$ dex and $8.15 \pm 0.10$ dex, employing the O values from \citet{Asplund2021} and \citet{Caffau2011}, respectively.

For Cl and Ar we adopt the values from \citet{Lodders2019}. These elements do not play any significant role in the calculation of the SSM, however, we include them for completeness.
\subsubsection{F, Na, K, P, Al, S}
We do not redetermine the abundance of these elements in this work, but rely on recent literature values, giving preference to the values that are most consistent with our methodological approach.

The solar abundance of F is taken from the analysis by \citet{Maiorca2014}. To the best of our knowledge, no NLTE analysis of F has been performed to date. 

For both Na and K, we adopt the estimates from \citet{Zhao2016}. They determined the Na abundance in NLTE using six Na I lines and the model atom from \cite{Gehren2004}. The error of the Na value adopted for this study was calculated from the line-to-line scatter (presented in Table~2 in \citealt{Zhao2016}). 
The study by \citet{Scott2015a} advocated a solar Na abundance of $6.21 \pm 0.04$ dex. 

The abundance of K determined by \citet{Zhao2016} and adopted for this study was based on the NLTE model presented in \citet{Zhang2006}. Another NLTE estimate of the solar K abundance by \citet{Reggiani2019}, A(K) $=5.11$ dex, is fully consistent with that of \citet{Zhao2016}.We adopt a conservative uncertainty of A(K) to be $0.1$ dex, as neither \citet{Zhao2016} nor \citet{Reggiani2019} provide an uncertainty on their K abundance estimate.

The abundance of P was taken from \citet{Scott2015a} and it is based on LTE, for the lack of NLTE calculations. The value of the Al abundance is taken from the 3D NLTE analysis by \citet{Nordlander2017}.

For S, we adopt the value by \citet{Caffau2007}, A(S) $=7.16 \pm 0.11$ dex. Their estimate is based on the detailed analysis of several S I lines in the optical and near-IR solar spectrum using the CO5BOLD model atmosphere using NLTE abundance corrections from \citet{Takeda2005}. Also the NLTE analyses by \citet{Korotin2009} and \citet{Korotin2017} suggest significant deviations from LTE for the majority of S I lines. Another recent estimate of the solar photospheric S abundance was proposed by \citet{Asplund2021}, and it is slightly lower (A(S) $=7.12 \pm 0.03$ dex) compared to our adopted value. 
\subsubsection{Fe-peak elements}
For Sc, we refer to the estimate provided by \citet{Zhao2016} using a set of Sc II lines modelled in NLTE with full account for hyperfine splitting (HFS). This estimate is preferred over other values, which were carried out in LTE \citep[e.g.][]{Lawler2019, Asplund2021}. It shall be noted that both Sc I and Sc II show significant departures from LTE, therefore especially the Sc II-based abundances are typically overestimated \citep{Zhao2016}. Our Sc value is in excellent agreement with the meteoritic abundance.

The solar abundances of Ti, Cr, and Co were adopted from \citet{Bergemann2011}, \citet{Bergemann2010b}, and \citet{Bergemann2010a}, respectively. All these estimates are based on detailed NLTE modelling, with account for HFS for Co lines and isotopic shifts for Ti.

Our value of the Mn abundance (A(Mn)$= 5.52 \pm 0.04$ dex) was taken from the detailed analysis in \citet{Bergemann2019}. This estimate is based on full 3D NLTE radiative transfer calculations of 13 Mn I lines, using the same 3D STAGGER model atmosphere as employed in this work. \citet{Bergemann2019} showed that taking into account 3D NLTE effects is essential to obtain reliable excitation-ionisation balance for Mn in atmospheric conditions of late-type stars.
\subsection{Meteoritic abundance scale \label{sec:meteo}} 

\begin{figure}[h!]
\hbox{
\includegraphics[width=0.99\linewidth]{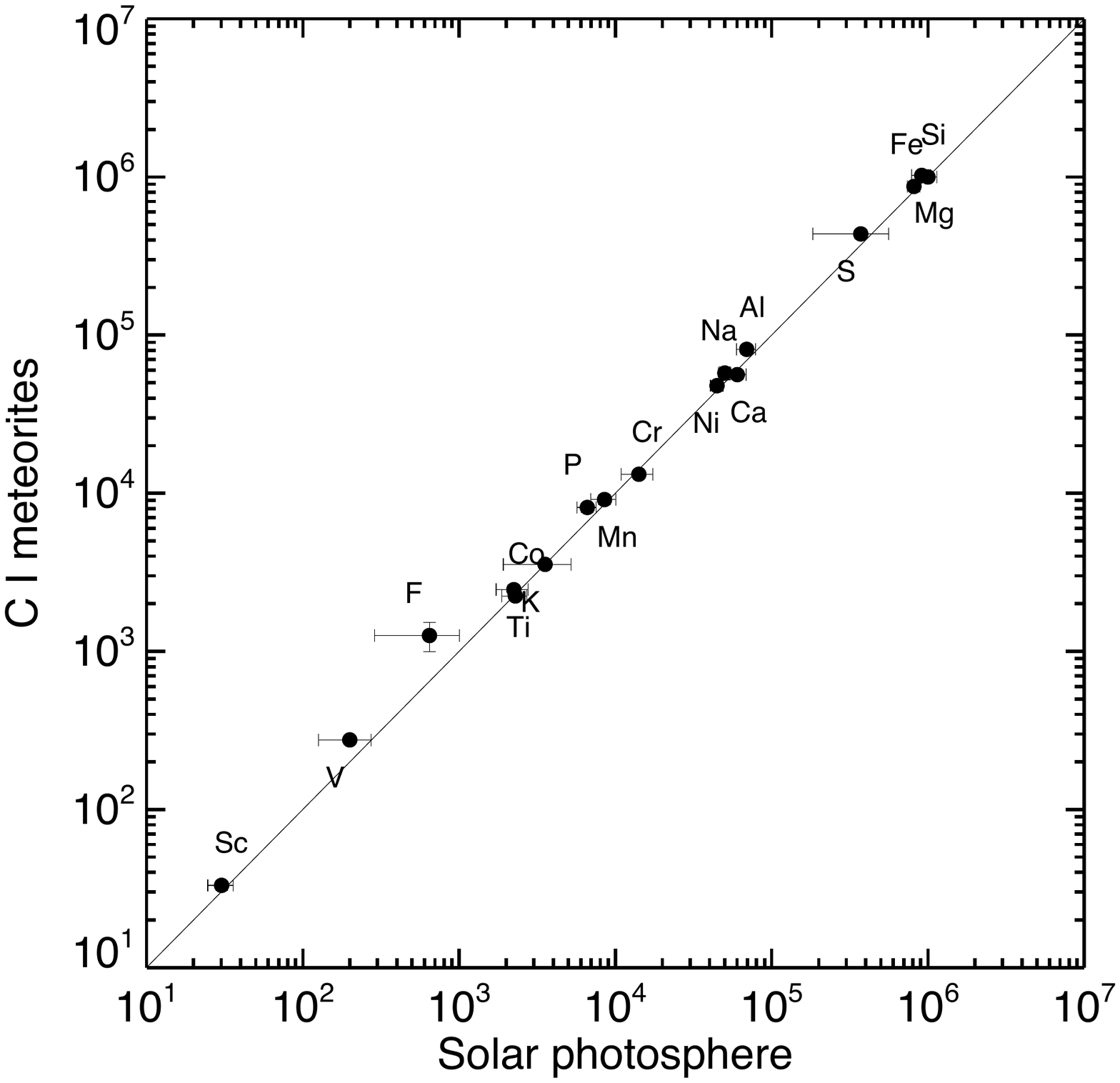}
}
\caption{Comparison of the photospheric estimates with meteoritic values.}
\label{fig:meteo}
\end{figure} 

CI-chondrites are a well known source to measure primitive solar system abundances of refractory elements \citep{Lodders2003,Lodders2021}. Meteoritic measurements have been historically more robust than photospheric determinations and the level of agreement between meteoritic and photospheric abundances has been used as a gauge of the quality of the latter. Bringing the meteoritic abundances to the photospheric scale requires defining an anchor point. Often, the Si abundance has been used for this purpose \citep{Grevesse1998,Lodders2003,Asplund2021}, but groups of elements have also been used \citep{Anders1989,Palme2014}.

Regardless of the methodology, meteoritic abundances converted to the photospheric scale have two uncertainty terms. The first one is the intrinsic error associated with the process of the abundance measurement that can be assumed uncorrelated among different elements. The second one is the systematic error associated with the transformation and it is a fully correlated error among all chemical elements. As a result, these two error sources cannot be quadratically added. This error source when meteoritic abundances are used for calculation of solar (stellar) models has been traditionally ignored and, unfortunately, can be equal or even larger than the measurement error. Our goal below is to obtain meteoritic solar abundances for which the systematic component of the error is minimized, i.e. the anchor point is defined more robustly. 

To this end, we adopt the CI-chondrite abundances from \cite{Lodders2021} in the cosmochemical scale, defined such that the Si abundance is $N({\rm Si})=10^6$. In order to derive meteoritic abundances in the astronomical scale we introduce a scale factor $c$. We determine the latter by minimizing the quadratic difference between the photospheric and meteoritic scales using the set of five refractory elements that have been newly derived in this work, constructed as
\begin{equation}
    \chi^2 = \sum_i{\frac{\left[A(i)_{\rm ph} - \left(\log{N_i}+c\right)\right]^2}{{\sigma_{\rm i,ph}^2 + \sigma_{\rm i,me}^2}}},
\end{equation}
where the sum extends over Mg, Si, Ca, Fe, and Ni. We obtain an excellent agreement between the two scales, yielding a total $\chi^2_{\rm min} = 1.27$, for $c=1.57\pm 0.02$\footnote{Before rounding off, the result is $c=1.567\pm0.023$.}. The uncertainty in $c$ is determined by the range over which $\Delta \chi^2=1$ around the minimum. It represents the systematic uncertainty associated to the meteoritic scale that should be included for all elements and is, in addition, a fully correlated error source among them. For comparison, \cite{Lodders2003} obtained $c=1.54$ using only Si, but did not include an estimate of the systematic error.
The final meteoritic abundances transformed to the solar photospheric abundance scale and the associated measurement errors are listed in Table~\ref{table:final-abundances}. The systematic and fully correlated 0.02~dex error is not included in the table but should always be taken into consideration when assessing the uncertainties due to chemical composition uncertainties in solar and stellar models.

Our final estimate of the solar $Z/X$ ratio is $0.0225$, if  calculated using the photospheric abundances only, and $0.0226$, if the meteoritic abundances are used for most species, except C,N, and O, for which the photospheric values are used. Our $Z/X$ ratio is $26\%$ higher compared to \citet{Asplund2009}, $Z/X = 0.0181$. It very close (within $1 \%$) to the photospheric and meteoritic estimates by \citet{Grevesse1998}, $Z/X = 0.0231$ and 0.0229) respectively, although the internal distribution of metals, i.e. the mixture, is different. And it is almost 10\% higher than  the estimate by \citet{Caffau2011}, $Z/X = 0.0209$.
\begin{figure}[h!]
\includegraphics[width=\columnwidth]{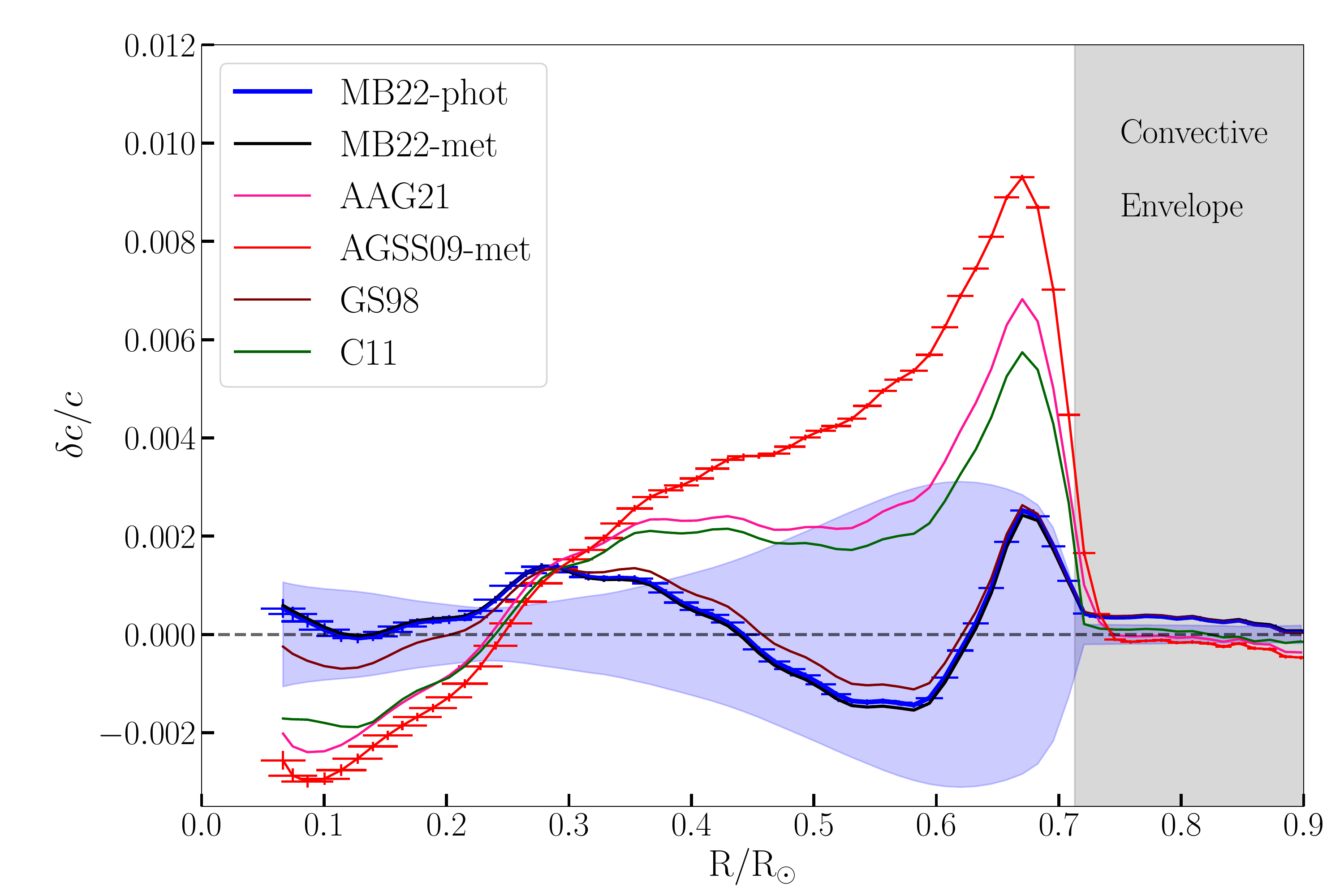}
\caption{The sound speed profiles of the SSM computed using different solar chemical mixtures, \citet[C11]{Caffau2011}, \citet[AGSS09-met]{Asplund2009}, \citet[GS98]{Grevesse1998}, \citet[AAG21]{Asplund2021}, and this work (MB22-phot and MB22-met, see Sect.~\ref{sec:meteo} for details. ). The shaded blue area represents the solar model uncertainties arising from the inputs to the model, see \cite{villante:2014} for details. Error bars in AGSS09-met and MB22-phot lines denote fractional sound speed uncertainties arising from helioseismic data (y-axis) and width of inversion kernels (x-axis).}
 \label{fig:soundspeed}
\end{figure} 

\subsection{Standard Solar Models}\label{sec:ssm}

The choice of the solar chemical mixture has a direct impact on SSMs because they are calibrated to reproduce the adopted photospheric chemical mixture at the present-day solar age. In this way, the abundance of metals in the interior of SSMs is determined by the photospheric abundances. Metals are main contributors to the radiative opacity in the solar interior which, in turn, determines the mechanical and thermal structure of the model. 

\begin{figure*}[h!]
\includegraphics[width=\columnwidth]{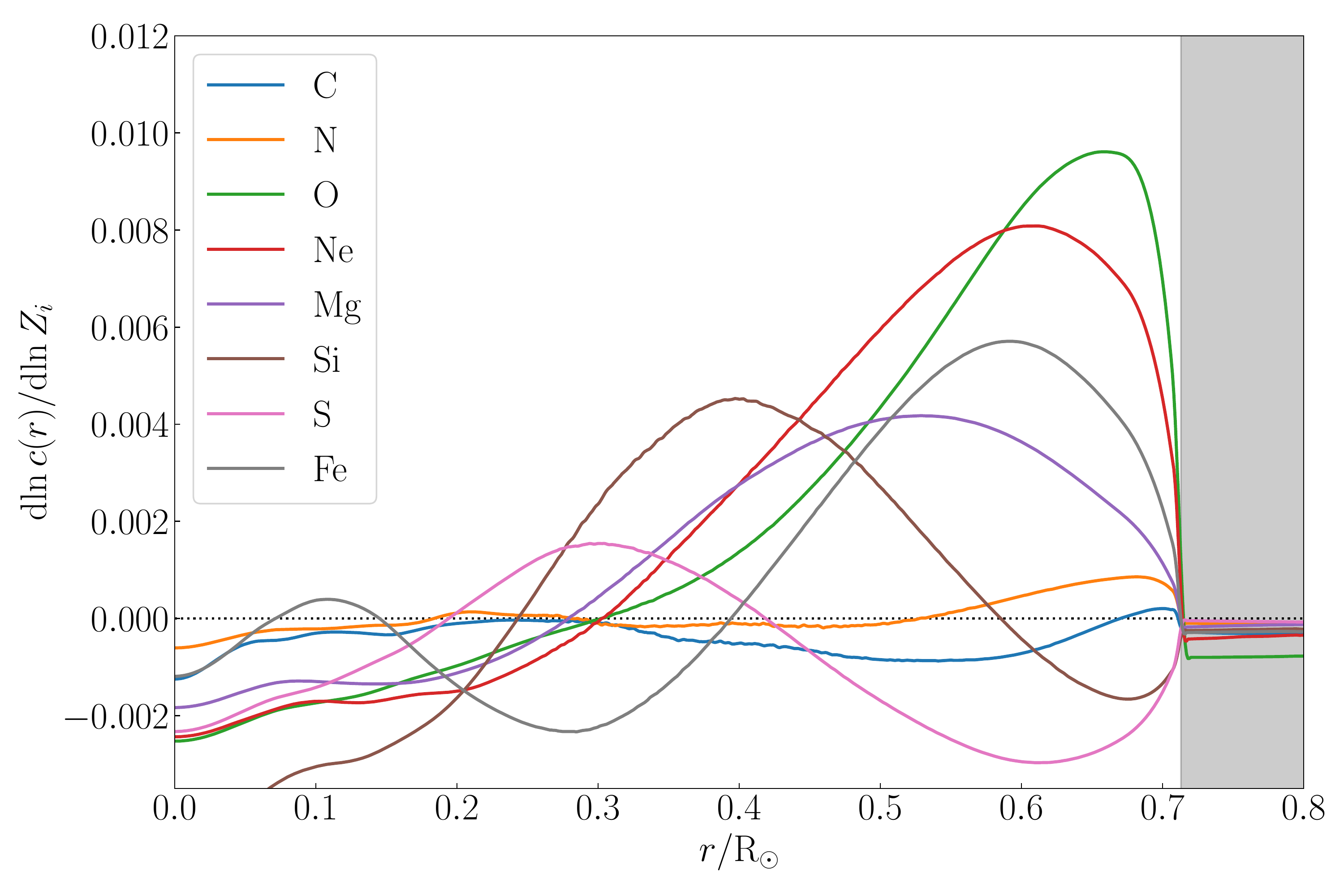}
\includegraphics[width=\columnwidth]{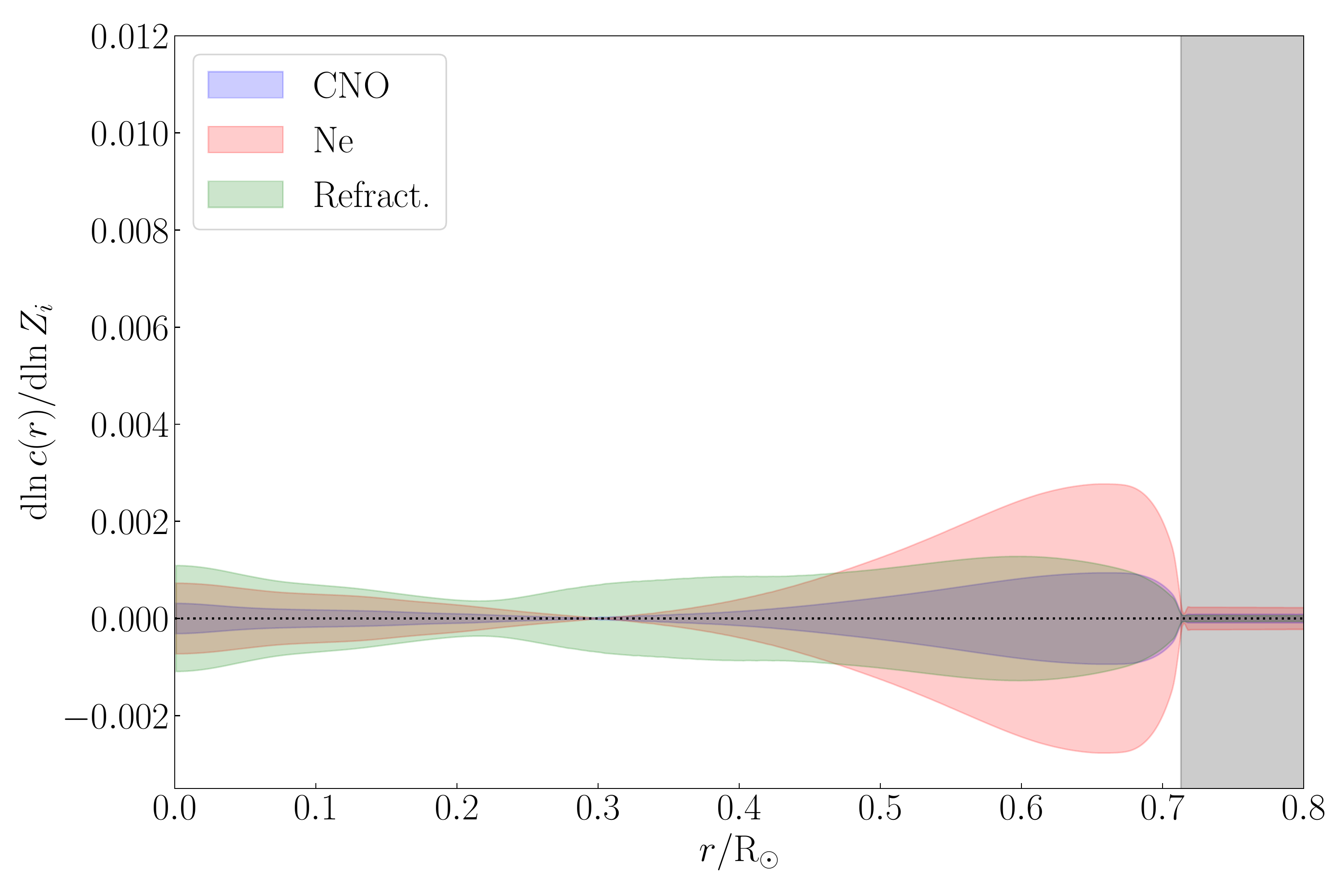}
\caption{Left panel: Logarithmic partial derivatives of the sound speed with respect to element abundances. Right panel: fractional $1\sigma$ uncertainties in the solar model sound speed based on photospheric abundances for different groups of elements as indicated.}
\label{fig:dcdz}
\end{figure*} 

A widely used diagnostic for the quality of SSMs is the comparison of the sound speed profile of the model against the solar profile as inferred from helioseismic techniques. The deficiency of SSMs calibrated on solar mixtures based on results from \citet{Asplund2005}, \citet{Asplund2009}, and the much better results for SSMs based on the solar mixtures by \citet{Grevesse1993, Grevesse1998} are well documented \citep{Bahcall2005, Basu2008, Serenelli2009, Pinsonneault2009, Serenelli2011, Buldgen2019}. The solar mixture by \citet{Caffau2011} slightly improves the comparison, owing to overall higher (10 to 30\%, depending on the element) abundances of CNO with respect to \citet{Asplund2009} and, to a minor extent larger abundances of refractories. 

\begin{table}[h!]
\renewcommand{\footnoterule}{} 
\setlength{\tabcolsep}{5pt}
\caption{Main characteristics of SSMs used in this work: depth of convective envelope ($R_{\rm CZ}$), surface helium mass fraction ($Y_{\rm S}$, fractional sound speed rms $\langle \delta c /c \rangle$ and initial helium $Y_{\rm ini}$ and metal $Z_{\rm ini}$ mass fractions.}
\label{table:ssm}
\begin{center}          
\begin{tabular}{lccccc}   
\hline\hline      
Model & $R_{\rm CZ}/{\rm R_\odot}$ & $Y_{\rm S}$ & $\langle \delta c /c \rangle$ & $Y_{\rm ini}$ & $Z_{\rm ini}$ \\ \hline
MB22-phot & 0.7123 & 0.2439 & 0.0010 & 0.2734 & 0.0176\\
MB22-met & 0.7120 & 0.2442 & 0.0010 & 0.2737 & 0.0178\\
AAG21 & 0.7197 & 0.2343 & 0.0027 & 0.2638 & 0.0155 \\
AGSS09-met & 0.7231 & 0.2316 & 0.0041 & 0.2614 & 0.0149 \\
GS98 & 0.7122 & 0.2425 & 0.0010 & 0.2718 & 0.0187 \\
C11 & 0.7162 & 0.2366 & 0.0021 & 0.2658 & 0.0169\\ 
\hline
\end{tabular}
\end{center}

\end{table}

Here, we have computed new SSMs using the \texttt{GARSTEC} code \citep{weiss:2008}, based on the photospheric and meteoritic solar mixtures provided in Table~\ref{table:final-abundances} (MB22-phot and MB22-met models respectively). The physics included in the models is the same as in \citet{Vinyoles:2017}. Atomic opacities have been computed for both flavors of MB22 solar mixtures, using OPCD 3.3 routines \citep{badnell:2005}. Opacities at low temperatures are most sensitive to the chemical composition thus we computed those using the revised MB22 mixture as described in \citet{ferguson:2005}. 

New helioseismic inversions of the solar sound speed have been performed for all the SSMs used in this work based on the methodology described in \citet{basu:2009} and references therein using the helioseismic data described in \citet{basu:2021}. The fractional sound speed differences are shown in Fig.~\ref{fig:soundspeed}. For comparison, we overplot the new inversion results for the SSMs computed with the \citet{Asplund2021} (AAG21), \citet{Caffau2011} (C11), \citet{Asplund2009} (AGSS09-met), and \citet{Grevesse1998} (GS98) compositions. The figure also includes, in shaded area, the uncertainty in the sound speed profile due to uncertainties in the SSM inputs computed using the method from \citet{villante:2014} (see also below).

Results show that SSMs based on the solar mixtures obtained in this work reproduce the solar sound speed profile with much higher accuracy than the AGSS09-met model, interestingly, at a comparable level to the GS98 SSM. This is due to the higher metal abundances found in this work, in particular for O, and to a lesser extent refractories, compared to \citet{Asplund2009} results.  The similarity between results of the MB22 SSMs and GS98 could be expected given that the global $(Z/X)_\odot$ is very similar. However, it is important to highlight that the internal distribution of elements is different in our and GS98 solar mixtures and the agreement in $Z/X$ is a numerical coincidence. For instance, for two critical elements, O and Si, our values are $15 \%$ lower and $10$\% higher, respectively, compared to GS98. For completeness, we have included a SSM based on the \citet{Asplund2021} composition which shows a partial improvement with respect to the AGSS09-met model, but still far from the SSMs based on the solar mixtures determined in the present work. 

To quantify the impact of solar abundances on the sound speed profile, we follow \citet{villante:2014} and show the logarithmic partial derivatives of the sound speed profile with respect to the abundance of individual elements in the left panel of Fig.~\ref{fig:dcdz}. These derivatives depend only slightly on the reference solar mixture but some differences with respect to \citet{villante:2014} can be observed. The right panel of Fig.~\ref{fig:dcdz} shows the actual fractional sound speed uncertainty in the models due to uncertainties is solar abundances, with chemical elements grouped as indicated in the figure and using the photospheric uncertainties given in Table~\ref{table:final-abundances}. The uncertainty in Ne plays a dominant role in the regions below the convective envelope. The uncertainty due to refractory elements is dominant below 0.45${\rm R_\odot}$. This is different from previous results on SSMs (see e.g. \citealt{Vinyoles:2017}) and is due to the larger spectroscopic uncertainties given in the present work.

A summary of relevant characteristics of SSMs used in this work is given in Table~\ref{table:ssm}. The depth of the convective envelope, $R_{\rm CZ}$, and the surface helium abundance $Y_{\rm S}$ are two widely used helioseismic probes of SSMs and should be compared to their observational values, $0.713 \pm 0.001{\rm R_\odot}$ \citep{basu:1997} and $0.2485\pm 0.0034$ \citep{basu:2004} respectively. Note that, despite the fact that our photospheric $Z/X$ is almost equal to that of GS98, $Z_{\rm ini}$ in the MB22 models is 5\% lower than in the GS98 SSM. This is due to the higher abundance of refractories in our results, which result in a slightly larger opacity in the solar core and a higher $Y_{\rm ini}$ (or lower initial hydrogen) to satisfy the solar luminosity constraint. In addition, we provide the averaged rms of the sound speed fractional difference. 
While it is beyond the scope of this paper to present a complete analysis of the comparison between SSMs and helioseismic observations, it is clear that our newly derived solar mixture improves the agreement between the observed helioseismic characteristics of the Sun and SSMs. Our results are close to those based on GS98 solar mixture, despite drastically different approaches.

As shown in Fig.~\ref{fig:meteo}, the agreement between the photospheric and meteoritic scales is excellent. This is also reflected by the similarity of the results obtained with the MB22-photo and MB22-met SSMs. 

The SSM presented in the current work provides a good consistency with the solar structure based on helioseismic observations. While SSMs offer an incomplete description of the physics in the solar interior, current results alleviate the need for more complex physics, such as accretion of metal-poor material \citep{Serenelli2011}, energy transport by dark matter particles \citep{Vincent2015}, revision of opacities \citep{Bailey2015}, enhanced gravitational settling and other effects \citep{guzik:2010}. 
\section{Conclusions}\label{sec:conclusions}

In this work, we used new observational material for the Sun, new updated atomic data, and up-to-date NLTE model atoms, to re-analyse the detailed chemical composition of the solar photosphere. For O I, we re-computed the oscillator strengths using several independent approaches, finding excellent agreement between the new values and those adopted in \citet{Bergemann2021}. New log(gf) values were also computed for N I transitions. We used two families of 3D radiation-hydrodynamics simulations of solar convection,  CO5BOLD and STAGGER, to represent the solar atmosphere, which allowed us for the first time to quantify the differences between the abundances inferred with both models. We focused on carrying out the analysis in such a way that it can be applied directly to stars observed within ongoing and upcoming large-scale spectroscopic surveys, such as 4MOST, WEAVE, and SDSS-V.

We provided new estimates of chemical abundances for elements most relevant for the calculations of standard solar models, including C, N, O, Mg, Si, Ca, Fe, and Ni. We complement these results with estimates of the solar abundances of Mn, Ti, Co, Cr, and Sr from our previous studies based on NLTE. Comparing our abundances of refractories with the element ratios based on CI chondrites, we find an excellent agreement between the two scales. We find that for most species our abundances are in good agreement with other literature values obtained with detailed NLTE methods, 1D and 3D model atmospheres \citep[e.g.][]{Caffau2011, Osorio2015, Alexeeva2018, Mashonkina2019}.

We determine the solar photospheric present-day $Z/X$ ratio of $0.0225$, when calculated using the photospheric abundances only, and $0.0226$, if the meteoritic abundances are used for most species, except C, N, and O, for which the photospheric values are used. Our estimates  are $26\%$ higher compared to those determined by \citet{Asplund2021}, but they are in a much better agreement with \citet{Caffau2011} and \citet{Grevesse1998}, the difference being $10\%$ and $1\%$, respectively. The very close numerical agreement of $Z/X$ with \citet{Grevesse1998} is, however, fortuitous, as abundances of individual elements are different in our and their study. Whereas the latter study made use of 1D LTE models, their uncertainties are more conservative (of the order 10 to 20 $\%$ for most elements) and appear to be more realistic, which accounts for the difference with \citet{Asplund2009} and \citet{Asplund2021}. 

Our detailed calculations of SSMs suggest that the presented in this study chemical composition leads to consistent results between the interior structure of the Sun and the helioseismic quantities that match results based on the old spectroscopic results (e.g. \citealt{Grevesse1993, Grevesse1998}). It is the first time that SSMs using  state-of-the-art spectroscopic results for solar abundances are able to reproduce the solar interior properties as determined through helioseismic techniques. We are confident this works brings us close to the solution of the \textit{solar abundance problem}, arisen in the early 2000s with the initial results on solar O based on 3D models and NLTE line formation \citep{Asplund2005}, a problem that had defied all attempted solutions in the form of non standard stellar physics. The residual differences can possibly be explained by other systematic limitations of stellar models \citep{Buldgen2019}.
\begin{acknowledgements}
We acknowledge support by the Collaborative Research centre SFB 881 (projects A4, A5, A10), Heidelberg University, of the Deutsche Forschungsgemeinschaft (DFG, German Research Foundation).  
MB is supported through the Lise Meitner grant from the Max Planck Society. 
AS is supported by the MICINN grant PRPPID2019-108709GB-I00 and the European Union’s Horizon 2020 research and innovation programme under grant agreement No 101008324 (ChETEC-INFRA) and by the Spanish program Unidad de Excelencia Maria de Maeztu CEX2020-001058-M.
BP is supported in part by The Centre  National d'Etudes Spatiales (CNES).
This project has received funding from the European Research Council (ERC) under the European Union’s Horizon 2020 research and innovation programme (Grant agreement No. 949173).
UH acknowledges support from the Swedish National Space Agency (SNSA/Rymdstyrelsen).
We thank Henrik Hartman for providing new atomic data for Si transitions.
EM would like to thank Richard Hoppe for many productive discussions.
We thank an anonymous referee for their helpful suggestions and comments.

\end{acknowledgements}

\newpage
\bibliographystyle{aa}
\bibliography{lit}
\label{lastpage}

\appendix
\section{Appendix} \label{sec:appendix}
Here we provide abundances derived from individual lines observed in the solar spectrum. For details on the methods see Sect.~\ref{sec:analysis}.
\begin{table}[h!]
\begin{minipage}{\linewidth}
\setlength{\tabcolsep}{15pt}
\caption{$\td$ NLTE abundances derived from individual spectral lines}
\label{tab:ind_abundances}     
\begin{center}
\begin{tabular}[t]{l c}
\noalign{\smallskip}\hline\noalign{\smallskip}  $\lambda [\AA]$ & A(el)  \\
\noalign{\smallskip}\hline\noalign{\smallskip} 
\ion{ C }{ i }  &  \\ 
 5052.160 & 8.61 \\ 
 6587.610 & 8.51 \\ 
 7113.180 & 8.57 \\ 
\ion{ N }{ i }  &  \\ 
 8629.200 & 8.05 \\ 
 8683.400 & 7.91 \\ 
\ion{ O }{ i }  &  \\ 
 6300.304 & 8.75 \\ 
 7771.940 & 8.76 \\ 
 7774.170 & 8.78 \\ 
 7775.390 & 8.79 \\ 
\ion{ Mg }{ i }  &  \\ 
 5528.405 & 7.51 \\ 
 5711.088 & 7.59 \\ 
\ion{ Si }{ i }  &  \\ 
 5645.600 & 7.60 \\ 
 5684.480 & 7.60 \\ 
 5690.425 & 7.62 \\ 
 5701.104 & 7.52 \\ 
 5772.146 & 7.62 \\ 
 5793.073 & 7.49 \\ 
 6741.640 & 7.71 \\ 
 7034.900 & 7.69 \\ 
 7226.210 & 7.58 \\ 
\ion{ Ca }{ i }  &  \\ 
 5260.387 & 6.30 \\ 
 5512.980 & 6.35 \\ 
 5867.562 & 6.42 \\ 
 6166.439 & 6.38 \\ 
 6455.598 & 6.37 \\ 
 6471.662 & 6.35 \\ 
 6499.650 & 6.39 \\ 
\end{tabular}
\begin{tabular}[t]{lc}
\noalign{\smallskip}\hline\noalign{\smallskip}  $\lambda [\AA]$ & A(El)  \\
\noalign{\smallskip}\hline\noalign{\smallskip} 
\ion{ Fe }{ i }  &  \\ 
 5242.491 & 7.52 \\ 
 5365.399 & 7.17 \\ 
 5379.574 & 7.55 \\ 
 5398.279 & 7.50 \\ 
 5560.212 & 7.61 \\ 
 5638.262 & 7.42 \\ 
 5661.346 & 7.44 \\ 
 5679.023 & 7.79 \\ 
 5731.762 & 7.69 \\ 
 5741.848 & 7.63 \\ 
 5855.077 & 7.49 \\ 
 5905.672 & 7.49 \\ 
 5930.180 & 7.63 \\ 
 6027.051 & 7.51 \\ 
 6056.005 & 7.41 \\ 
 6093.644 & 7.65 \\ 
 6165.360 & 7.56 \\ 
 6187.990 & 7.53 \\ 
 6270.225 & 7.47 \\ 
\ion{ Fe }{ ii }  &  \\ 
 5234.625 & 7.39 \\ 
 5325.553 & 7.43 \\ 
 5425.257 & 7.45 \\ 
 5543.936 & 7.55 \\ 
 6084.111 & 7.59 \\ 
 6456.383 & 7.54 \\ 
\ion{ Ni }{ i }  &  \\ 
 4740.170 & 6.26 \\ 
 4811.980 & 6.26 \\ 
 4814.600 & 6.19 \\ 
 4976.130 & 6.20 \\ 
 5157.980 & 6.13 \\ 
 5537.100 & 6.07 \\ 
 6176.820 & 6.29 \\ 
 6204.600 & 6.24 \\ 
 6223.990 & 6.23 \\ 
 6414.590 & 6.24 \\ 
\ion{ Ni }{ ii }  &  \\ 
 6378.260 & 6.26 \\ 
 \noalign{\smallskip}\hline\noalign{\smallskip}
 \end{tabular}
\end{center}
\end{minipage}
\end{table}
\end{document}